\documentclass[nofootinbib,showpacs,preprint,preprintnumbers,amsmath,amssymb]{revtex4-1}

\usepackage{graphicx}
\usepackage{dcolumn}
\usepackage{bm}
\usepackage{amsmath}
\usepackage{mathrsfs}
\usepackage{multirow}
\usepackage{xcolor}
\usepackage{enumitem}

\newcommand{\gsim}{\gtrsim}
\newcommand{\lsim}{\lesssim}

\newcommand{\lf}{\left(}
\newcommand{\ri}{\right)}

\newcommand{\nn}{\nonumber}

\newcommand{\sqs}{\sqrt{s}}

\renewcommand{\lg}{\mathscr{L}}

\newcommand{\mcd}{\mathcal{D}}
\newcommand{\mcf}{\mathcal{F}}

\newcommand{\mco}{\mathcal{O}}
\newcommand{\mcq}{\mathcal{Q}}

\newcommand{\mcu}{\mathcal{U}}

\newcommand{\bmcf}{\bar{\mathcal{F}}}
\newcommand{\bmcq}{\overline{\mathcal{Q}}}
\newcommand{\bmcu}{\overline{\mathcal{U}}}
\newcommand{\bmcd}{\overline{\mathcal{D}}}

\newcommand{\br}{\mathcal{B}}
\newcommand{\hc}{{\rm H.c.}}

\newcommand{\pb}{{~{\rm pb}}}
\newcommand{\fb}{{~{\rm fb}}}

\newcommand{\gev}{{~{\rm GeV}}}
\newcommand{\tev}{{~{\rm TeV}}}
\def\gammu{\gamma^\mu}

\newcommand{\beq}{\begin{equation}}
\newcommand{\eeq}{\end{equation}}
\newcommand{\bea}{\begin{eqnarray}}
\newcommand{\eea}{\end{eqnarray}}
\newcommand{\barr}{\begin{array}}
\newcommand{\earr}{\end{array}}
\newcommand{\bc}{\begin{center}}
\newcommand{\ec}{\end{center}}
\newcommand{\bit}{\begin{itemize}}
\newcommand{\eit}{\end{itemize}}
\newcommand{\ben}{\begin{enumerate}}
\newcommand{\een}{\end{enumerate}}

\newcommand{\al}{\alpha}
\newcommand{\bt}{\beta}

\newcommand{\dt}{\delta}
\newcommand{\Dt}{\Delta}

\newcommand{\sg}{\sigma}
\newcommand{\es}{\epsilon}
\newcommand{\kp}{\kappa}

\newcommand{\gm}{\gamma}
\newcommand{\Gm}{\Gamma}

\newcommand{\mch}{M_{H^\pm}}

\newcommand{\ca}{c_\alpha}
\newcommand{\sa}{s_\alpha}

\newcommand{\tb}{t_\beta}
\newcommand{\cb}{c_\beta}
\renewcommand{\sb}{s_\beta}

\newcommand{\cba}{c_{\beta-\alpha}}
\newcommand{\sba}{s_{\beta-\alpha}}

\newcommand{\gh}{\hat{g}}

\newcommand{\mh}{m_{h}}
\newcommand{\mhh}{M_{H}}

\newcommand{\ttop}      {{t\bar{t}}}

\newcommand{\bb}      {{b \bar{b}}}

\def\gev{\,{\rm GeV}}
\def\tev{\,{\rm TeV}}
\def\pb{\,{\rm pb}}

\def\nn{\nonumber}
\def\gammu{\gamma^\mu}
\def\tb{t_\beta}

\def\br{{\rm Br}}

\begin{document}

\baselineskip 3.5ex
\vspace*{18pt}


\title{The $W \gamma$ decay of the elusive charged Higgs boson
\\
in the two-Higgs-doublet model with vectorlike fermions}

\author{Jeonghyeon Song, and Yeo Woong Yoon}
\affiliation{Department of Physics, Konkuk University, Seoul 05029, Korea}


\begin{abstract}
The LHC search for the charged Higgs boson in the intermediate-mass range
($M_{H^\pm} \sim m_t$) is actively being
performed after the next-to-leading order calculation of the total production cross section of $p p \to H^\pm W^\mp b \bar{b}$.
In the decay part, only the $\tau\nu$ mode is mainly concerned
because of the experimental difficulty in the $t b$ mode.
In the framework of a two-Higgs-doublet model,
we suggest that the $W \gamma$ channel
can be helpful
in probing this charged Higgs boson, 
if introducing vectorlike fermions.
In type-I-II model where the SM fermions are assigned in type-I while the vectorlike fermions are in type-II,
the branching ratio is greatly enhanced
up to $\sim \mathcal{O}(0.01)$
in a large portion of the parameter space allowed by
the Higgs precision data, the electroweak oblique parameters, and the direct search bounds at the LHC.
Two kinds of production channels, $gg\to H^\pm W^\mp b \bar{b}$ and $g g \to H/A \to H^\pm W^\mp$,
are also studied.
We find that the signal rate $\sigma \times \mathcal{B}(W \gamma)$
is quite sizable, more than 10 fb in some parameter space.
\end{abstract}

\maketitle

\section{Introduction}

The repeated phrase ``no excess above the estimated standard model (SM) background"
in every new physics study at the LHC
is disappointing to many particle physicists.
Before we are resigned to no prospect of the new signal,
we need to draw upon the planned high-luminosity of the LHC.
We shall be able to probe some faint signals if any. 
In the meantime, all we can and should do
is to search every hole and corner of the given parameter space.
The common method of finding a new particle is to
consider the main production channel and the main decay mode,
which spans the bulk of the parameter space most effectively. 
In order to target faint signals,
however,
each portion of parameter space
requires a customized approach, which is sometimes unorthodox. 

A good example is a charged Higgs boson $H^\pm$ 
in the two-Higgs-doublet model (2HDM)~\cite{Branco:2011iw}.
Since the 
$H^\pm$-$W^\mp$-$(Z/\gamma)$
vertex does not exist at the tree level
and the $H^\pm$-$W^\mp$-$h$  vertex  vanishes in the alignment limit~\cite{Campos:2017dgc},
the charged Higgs boson mostly decays into fermions in the normal scenario of $\mh=125\gev < M_A$ \footnote{
In the inverted scenario where the observed Higgs boson is the heavy CP-even $H$ ($\mhh=125\gev$),
the decay of $H^\pm \to W^\pm h$ is dominant in the alignment limit of $\cba=1$~\cite{Arbey:2017gmh}.
}.
The search strategy at the LHC~\cite{Aad:2014kga,Khachatryan:2015qxa,Aad:2013hla,Khachatryan:2015uua,Sirunyan:2018dvm,Aaboud:2016dig,Aad:2015typ}
depends on the charged Higgs boson mass $\mch$.
If it is significantly lighter than the top quark mass ($\mch \lsim 140\gev$), 
$H^\pm$ is mainly produced from the 
on-shell top quark decay
in the top quark pair production,
and then decays into $\tau\nu$.
If it is heavy like $\mch \gsim 200\gev$,
the production channel is $g \bar{b} \to H^+ \bar{t}$, followed by the decay $H^+ \to t \bar{b}$.

 As for the intermediate-mass region with $140\gev \lsim \mch \lsim 200\gev$, the interplay between the top quark resonant and non-resonant diagrams is important. 
Recently, 
the total production cross section 
of the full process  $pp \to H^\pm W^\mp \bb$ including resonant and non-resonant top quark diagrams was computed at  next-to-leading order (NLO)  accruacy~\cite{Degrande:2016hyf}.
This NLO computation stimulates the charged Higgs search for the intermediate mass region. 
In the decay part, only the $\tau\nu$ mode has been searched for~\cite{Aaboud:2018gjj,Sirunyan:2019hkq},
because the $tb$ mode is experimentally very difficult. 
Being kinematically below the $tb$ threshold,
the intermediate-mass charged Higgs boson has more chance to be probed through other decay channels 
since the $tb$ decay channel, once open, is severely dominant. 
Therefore, alternative decay modes shall be especially helpful for the intermediate-mass charged Higgs boson.

Non-fermionic decay channels of $H^\pm$ 
are only
the radiative decays of $H^\pm \to W^\pm \gm$ and $H^\pm \to W^\pm Z$.
The $W \gm$ and $W Z$ modes as a new resonance search at the LHC~\cite{Aad:2014fha,Aaboud:2018fgi} have been studied
in other new physics models.
A representative one is the Georgi-Machacek (GM) model~\cite{Georgi:1985nv}
where the custodial-fiveplet (both singly and doubly) charged  Higgs boson
is fermiphobic, mainly decaying into $WZ$ or $WW$ through the tree level couplings~\cite{Chiang:2012cn,Chiang:2018cgb,Hartling:2014zca,Chiang:2014bia}.
Below the kinematic threshold, 
the loop-induced decays into
$W\gm$
 were studied~\cite{Logan:2018wtm,Degrande:2017naf}.
In a generalized inert doublet model with a broken $Z_2$ symmetry,
called the stealth Higgs doublet model,
$H^\pm \to W^\pm\gm$ was also studied~\cite{Enberg:2013jba}.

In the usual 2HDM,  the branching ratios of both $W\gm$ and $WZ$ modes
are very suppressed, at most $\sim \mco(10^{-4})$.
So we question whether
the branching ratios can be meaningfully enhanced
if we extend the 2HDM by introducing
vectorlike fermions (VLFs)~\cite{Ellis:2015oso}.
A VLF with a mass around the electroweak scale
appears in many new physics models~\cite{Lavoura:1992np,Aguilar-Saavedra:2013qpa}.
One of the biggest advantages of VLFs is the consistency
with the Higgs precision data unlike heavy chiral fermions~\cite{Anastasiou:2011qw,Anastasiou:2016cez}.
However, enhancing the branching ratios of the radiative decays
is very challenging.
Naively raising the Yukawa couplings of the VLFs with the charged Higgs boson
shall confront
the constraints from the electroweak oblique parameters
since the VLF loop corrections to the vertex of $H^\pm$-$W^\mp$-$V$ ($V=\gm,Z$)
are usually correlated with those to the vacuum polarization amplitudes
of the SM gauge bosons.
We need to contrive a model which accommodates
significantly large loop-induced decays 
while satisfying the other direct and indirect constraints.
As shall be shown,
if we assign the SM fermions in type-I and the new VLFs in type-II,
the goal is achieved.
In a large portion of the parameter space,
$\br(H^\pm \to W^\pm \gm)$ for the intermediate-mass charged Higgs boson
is greatly enhanced by a few orders of magnitude.
However, the $WZ$ decay mode does not change much
because of the strong correlation with the electroweak oblique parameter $\hat T$.
This is our main result.

The paper is organized in the following way.
In Sec.~\ref{sec:review},
we review our model, the 2HDM with the SM fermions in type-I
and the VLFs in type-II.
Section \ref{sec:constraints} deals with indirect and direct constraints
such as
the Higgs precision data,
the direct searches for the charged Higgs boson and the VLFs at the LHC,
and
the electroweak oblique parameters.
Particularly for the electroweak oblique parameter $\hat T$,
we shall suggest our ansatz for the parameters.
In Sec.~\ref{sec:calculation},
we first present the one-loop level calculation of the decay rates of $H^\pm \to W^\pm \gm/W^\pm Z$
via the VLF loops.
This is a new calculation.
Then, we show that the branching ratio of $H^\pm \to W^\pm \gm$
can be highly enhanced by one or two orders of magnitude,
relative to that without the VLF contributions.
Section \ref{sec:production} covers the production channels of the charged Higgs boson in our model
as well as the 13 TeV LHC sensitivity to the $H^\pm \to W^\pm\gm$ mode.
Section \ref{sec:conclusions} contains our conclusions.

\newpage 

\section{2HDM with Vectorlike Fermions}
\label{sec:review}

We consider a 2HDM with vectorlike fermions in the alignment limit.
The Higgs sector is extended by introducing
two complex $SU(2)_L$ Higgs doublet scalar fields, $\Phi_1$ and $\Phi_2$~\cite{Branco:2011iw}:
\bea
\label{eq:phi:fields}
\Phi_i = \left( \begin{array}{c} w_i^+ \\[3pt]
\dfrac{v_i +  h_i + i \eta_i }{ \sqrt{2}}
\end{array} \right),
\eea
where $i=1,2$, and $v_{1,2}$ are the nonzero vacuum expectation values (VEVs) of $\Phi_{1,2}$.
Using the simplified notation of $s_x=\sin x$, $c_x = \cos x$, and $t_x = \tan x$, 
we take $\tb =v_2/v_1$.
The electroweak symmetry breaking occurs by the nonzero VEV of $v =\sqrt{v_1^2+v_2^2}=246\gev $.

The fermion sector of the SM is also extended by introducing
one $SU(2)$ doublet VLF and two $SU(2)$ singlet VLFs as follows:
\begin{eqnarray}
&&{\rm VLF~doublets:}~~\, \mcq_L = \left(\begin{array}{c} \mcu'_L \\ \mcd'_L \end{array} \right)
,~\mcq_R = \left(\begin{array}{c} \mcu'_R \\ \mcd'_R \end{array} \right)\,, \nn \\
&&{\rm VLF~singlets:}
~~~ \mcu_L,\quad \mcu_R, \quad \mcd_L , \quad \mcd_R \,.
\end{eqnarray}
Here $\mcu^{(\prime)}$ and $\mcd^{(\prime)}$
denote the up-type and down-type fermions, respectively.
We shall consider various kinds of the VLFs:
$(X,T)$, the vectorlike quark (VLQ) with the electric charges of $(5/3,2/3)$;
$(T,B)$, the VLQ with $(2/3,-1/3)$;
$(B,Y)$, the VLQ with $(-1/3,-4/3)$;
$(N,E)$, the vectorlike lepton (VLL)  with  $(0,-1)$~\cite{Aguilar-Saavedra:2013qpa}.

In order to avoid the flavor changing neutral currents (FCNC) at tree level,
we introduce a discrete $Z_2$ symmetry under which $\Phi_1 \to \Phi_1$
and $\Phi_2 \to -\Phi_2$~\cite{Glashow:1976nt,Paschos:1976ay}.
The  $Z_2$ parities of $\Phi_1$ and $\Phi_2$ dictate the scalar potential to be
\bea
\label{eq:V}
V_\Phi = && m^2 _{11} \Phi^\dagger _1 \Phi_1 + m^2 _{22} \Phi^\dagger _2 \Phi_2
-m^2 _{12} ( \Phi^\dagger _1 \Phi_2 + \hc) \nn \\
&& + \frac{1}{2}\lambda_1 (\Phi^\dagger _1 \Phi_1)^2
+ \frac{1}{2}\lambda_2 (\Phi^\dagger _2 \Phi_2 )^2
+ \lambda_3 (\Phi^\dagger _1 \Phi_1) (\Phi^\dagger _2 \Phi_2)
+ \lambda_4 (\Phi^\dagger_1 \Phi_2 ) (\Phi^\dagger _2 \Phi_1) \nn\\
&& + \frac{1}{2} \lambda_5
\left[
(\Phi^\dagger _1 \Phi_2 )^2 +  {\rm h.c.}
\right] , \label{eq:potential}
\eea
where we allow softly broken $Z_2$ parity but maintain the \textit{CP} invariance.
Five physical Higgs bosons (the light \textit{CP}-even scalar $h$ at a mass of 125 GeV,
the heavy \textit{CP}-even scalar $H$, the \textit{CP}-odd pseudoscalar $A$,
and two charged Higgs bosons $H^\pm$)
are related with the weak eigenstates via
\bea
\left(
\begin{array}{c}
h_1 \\ h_2
\end{array}
\right) =
\mathbb{R}(\al)
\left(
\begin{array}{c}
H \\ h
\end{array}
\right),
\quad
\left(
\begin{array}{c}
\eta_1 \\ \eta_2
\end{array}
\right) =
\mathbb{R}(\beta)
\left(
\begin{array}{c}
z^0 \\ A
\end{array}
\right)
, \quad
\left(
\begin{array}{c}
w_1^\pm \\ w_2^\pm
\end{array}
\right) =
\mathbb{R}(\bt)
\left(
\begin{array}{c}
w^\pm \\ H^\pm
\end{array}
\right),
\eea
where $z^0$ and $w^\pm$ are the Goldstone bosons that will be eaten by the $Z$ and $W$ bosons, respectively.
The rotation matrix $\mathbb{R}(\theta)$ is
\bea
\mathbb{R}(\theta) = \left(
\begin{array}{cr}
c_\theta & -s_\theta \\ s_\theta & c_\theta
\end{array}
\right).
\eea
The SM Higgs field is a linear combination of $h$ and $H$,
$
h_{\rm SM} = \sba h + \cba H$.
Because the observed Higgs boson at a mass of 125 GeV is very SM-like,
we take
the alignment limit of
\bea
\label{eq:alignment}
&& \bt-\al = \frac{\pi}{2} \quad\hbox{ (alignment limit) }.
\eea

\begin{table}
\begin{center}
  {\renewcommand{\arraystretch}{1.2}
\begin{tabular}{|c||c|c|c|}
\hline
~~~~~SM~~~~~& ~~~$Q_L,~L_L$~~~ & ~~~$u_R$~~~ & ~~~$d_R,~\ell_R$~~~ \\ \hline
type-I & $+$ & $-$ & $-$ \\ \hline \hline
VLF & $\mcq_{L,R}$ & $\mcu_{L,R}$ & $\mcd_{L,R}$ \\ \hline
type-II & $+$ & $-$ & $+$ \\ \hline
\end{tabular}
}
\caption{\label{tab:Z2}
The $Z_2$ parities of the SM fermions and VLFs.
}
\end{center}
\end{table}

The fermions can have different $Z_2$ parities.
For the SM fermions, we fix $Q_L \to Q_L$ and $L_L \to L_L$ under $Z_2$ parity transformation.
Then, there are
four different choices of $Z_2$ parities for the right-handed SM fermion fields,
leading to type-I, type-II, type-X, and type-Y.
The VLFs
need not to have the same $Z_2$ parity with the SM fermions.
Since our main purpose is to explore the possibility of highly enhancing $\br(H^\pm \to W^\pm \gm/W^\pm Z^0)$,
we consider type-I-II, where the SM fermions are assigned in type-I while the VLFs are in type-II
(see Table \ref{tab:Z2}).
The Lagrangian for the mass and Yukawa terms of the VLFs is then
\bea
\label{eq:L:Yuk:0}
- {\cal L}_{\rm Yuk} &=& M_\mcq \overline{\mcq} \mcq + M_\mcu \overline{\mcu} \mcu + M_\mcd \overline{\mcd}\mcd
+ \Big[   Y_{\mcd} \bmcq \Phi_1 \mcd+
 Y_{\mcu} \bmcq \,\widetilde{\Phi}_2 \mcu
 + {\rm h.c.}
  \Big] \,,
\eea
where $\widetilde{\Phi}= i \tau_2 \Phi^*$
and we assume
$Y_{\mcu}^L = Y_{\mcu}^R\equiv Y_{\mcu}$ and
$Y_{\mcd}^L = Y_{\mcd}^R\equiv Y_{\mcd}$.

The VLF masses are from the Dirac mass parameters as well as
from the Higgs VEVs.
The mass matrices ${\mathbb M_\mcd}$ and ${\mathbb M_\mcu}$
in the basis of $(\mcd',\mcd)$ and $(\mcu',\mcu)$, respectively,  are
\begin{eqnarray}
{\mathbb M_\mcd} = \left(\begin{array}{cc} M_\mcq  & \tfrac{1}{\sqrt{2}} Y_{\mcd} v \cb  \\
\tfrac{1}{\sqrt{2}}Y_{\mcd} v \cb  & M_\mcd \end{array} \right),
\quad
{\mathbb M_\mcu} = \left(\begin{array}{cc} M_\mcq  & \tfrac{1}{\sqrt{2}} Y_{\mcu} v \sb  \\
\tfrac{1}{\sqrt{2}}Y_{\mcu} v \sb  & M_\mcu \end{array} \right).
\end{eqnarray}
In the large $\tb$ limit where $\cb\ll1$ and $\sb \approx 1$,
the off-diagonal terms of
${\mathbb M_\mcd}$ are suppressed.
The VLF mass matrices are diagonalized by
the rotation matrices $\mathbb{R}(\theta_{\mcf})$ as
$ \mathbb{R}(\theta_{\mcf}) {\mathbb M_\mcf} \mathbb{R}^T(\theta_{\mcf}) = {\rm diag}(M_{\mcf_1},M_{\mcf_2})$
for $\mcf=\mcu,\mcd$,
leading to the mass eigenstates of the VLFs as
\begin{eqnarray}
\left(\begin{array}{c} \mcd_{1} \\ \mcd_{2} \end{array} \right) =
\mathbb{R}(\theta_{\mcd}) \left(\begin{array}{c} \mcd' \\ \mcd \end{array} \right),~~
\left(\begin{array}{c} \mcu_{1} \\ \mcu_{2} \end{array} \right) =
\mathbb{R}(\theta_{\mcu}) \left(\begin{array}{c} \mcu' \\ \mcu \end{array} \right)\,.
\end{eqnarray}
When $\theta_{\mcu,\mcd} \ll 1$, $\mcu_1$ and $\mcd_1$ are $SU(2)$ doublet-like while
$\mcu_2$ and $\mcd_2$ are singlet-like.
In what follows, we use $s_{\mcu}  = s_{\theta_{\mcu}} $
and $c_{\mcu}  = c_{\theta_{\mcu}} $ for notational simplicity.
The VLF mixing angles satisfy
\begin{equation}
\label{eq:mixingangle}
s_{2\mcd} = \frac{\sqrt{2}Y_\mcd v}{M_{\mcd_2}-M_{\mcd_1}}\cb\,,
\quad
s_{2\mcu} = \frac{\sqrt{2}Y_\mcu v}{M_{\mcu_2}-M_{\mcu_1}}\sb\,.
\end{equation}

The Yukawa Lagrangian for the SM fermions and the VLFs 
in terms of mass eigenstates is
\begin{eqnarray}
-{\cal L}_{\rm Yuk} &=&
\sum_{f=t,b,\tau} \frac{m_f}{v}
\lf \kp_f \bar{f}f h +  \xi^H_f \bar{f}f H  - i \xi^A_f \bar{f} \gm_5 f A \ri
\\ \nn &&
-
\left\{
\frac{\sqrt{2}}{v}\bar{t}
\lf
m_t \xi^A_t P_L + m_b \xi^A_b P_R
\ri b H^+
+\frac{\sqrt{2}}{v} m_\tau \xi_\ell^A \bar{\nu}_L \tau_R H^+
+{\rm h.c.}
\right\}
\\ \nn &&
+\sum_{\mcf}\sum_{i,j}
\sum_{\phi}y^{\phi}_{ \mcf_i \mcf_j } \, \phi \bmcf_i \mcf_j
+\sum_{\mcf}\sum_{i,j}
\left[ -i y^{A}_{ \mcf_i \mcf_j }
A
\bmcf_{i,R} \mcf_{j,L}
+ {\rm h.c.} \right]
 \\ \nn &&
  +\sum_{i,j} \Big[
y^{H^+}_{\mcu_i \mcd_j}  H^+ \bmcu_i \mcd_j
+{\rm h.c.} \Big],
\end{eqnarray}
where $\mcf={\mcu,\mcd}$, $i,j=1,2$, and $\phi=h,H$.
In our  type-I-II model, the normalized Yukawa couplings are
 \bea
 \label{eq:kp:xi}
 \kp_f &=& 1, \quad
 \xi^H_f = \frac{s_\alpha}{s_\beta},
 \quad
 \xi^A_u = -\xi^A_d = -\xi^A_\ell = \frac{1}{\tb},
\\ \nn
\xi^h_\mcd &=& -\sa,\quad \xi^h_\mcu =\ca,
\quad \xi^H_\mcd =\ca, \quad
\xi^H_\mcu= \sa,
\quad
\xi^A_\mcd=\sb,\quad \xi^A_\mcu = \cb.
 \eea
 Additionally, we shall impose the alignment condition, $\sba=1$.

The Yukawa couplings of the VLFs with neutral Higgs bosons
are
\begin{eqnarray}
\label{eq:Yukawa:coupling:m:eigen}
y^{\phi}_{\mcf_1 \mcf_1} &=& - y^\phi_{\mcf_2 \mcf_2} = -\frac{1}{\sqrt{2}}\,Y_\mcf \xi^{\phi}_\mcf s_{2\mcf} ,
\\ \nn
y^{\phi}_{\mcf_1 \mcf_2} &=&  y^{\phi}_{\mcf_2 \mcf_1} =\frac{1}{\sqrt{2}}\, Y_\mcf \xi^{\phi}_\mcf c_{2\mcf},
\\ \nn
y^{A}_{\mcf_i \mcf_i} &=&0,
\\ \nn
y^A_{\mcf_1 \mcf_2} &=& - y^A_{\mcf_2 \mcf_1} = \frac{1}{\sqrt{2}}\, Y_\mcf \xi^A_\mcf,
\end{eqnarray}
and those with the charged Higgs boson are
\begin{eqnarray}
\label{eq:cH:Yukawa:coupling:m:eigen}
y^{H^+}_{\mcu_1 \mcd_1}&=& \phantom{-} Y_\mcu \xi^A_\mcu c_\mcd s_\mcu + Y_\mcd \xi^A_\mcd s_\mcd c_\mcu\,,\\ \nn
y^{H^+}_{\mcu_1 \mcd_2} &=& \phantom{-} Y_\mcu \xi^A_\mcu s_\mcd s_\mcu - Y_\mcd \xi^A_\mcd c_\mcd c_\mcu\,,\\ \nn
y^{H^+}_{\mcu_2 \mcd_1} &=& - Y_\mcu \xi^A_\mcu c_\mcd c_\mcu + Y_\mcd \xi^A_\mcd s_\mcd s_\mcu \,,\\ \nn
y^{H^+}_{\mcu_2 \mcd_2} &=&  -Y_\mcu \xi^A_\mcu s_\mcd c_\mcu - Y_\mcd \xi^A_\mcd c_\mcd s_\mcu\,.
\end{eqnarray}

The gauge interaction Lagrangian of the VLF mass eigenstates is
\begin{eqnarray}
{\cal L}_{\rm gauge} &=& \sum_\mcf
\left[ e A_\mu \sum_{i}  Q_\mcf \bmcf_i \gammu \mcf_i
+ g_Z Z_\mu \sum_{i,j} \hat{g}^Z_{\mcf_i\mcf_j} \bmcf_i \gm^\mu \mcf_j
\right]
\\ \nn &&
+\frac{g}{\sqrt{2}}
\lf \hat{g}^W_{\mcd_i \mcu_j}   W_\mu^- \bmcd_i \gammu \mcu_j + {\rm h.c.} \ri
\,,
\end{eqnarray}
where $\mcf=\mcu,\mcd$, $g_Z =g/c_W$, and $c_W$ is the cosine of the electroweak mixing angle.
The normalized gauge couplings are
\begin{eqnarray}
\label{eq:ghat}
&& \hat{g}^Z_{\mcf_1 \mcf_1} = g_V^{\mcf'} c_{\mcf}^2 + g_V^{\mcf} s_\mcf^2\,,
\quad
\hat{g}^Z_{\mcf_2 \mcf_2} =
g_V^{\mcf'} s_\mcf^2 + g_V^\mcf c_\mcf^2\,, \\[3pt] \nn
&& \hat{g}^Z_{\mcf_1 \mcf_2} =\hat{g}^Z_{\mcf_2 \mcf_1} =
(g_V^{\mcf'}-g_V^\mcf)s_\mcf c_\mcf,~ \\[3pt] \nn
&&\hat{g}^{W}_{\mcd_1 \mcu_1} =  c_\mcu c_\mcd\,, \quad
\hat{g}^{W}_{\mcd_1 \mcu_2} =  s_\mcu c_\mcd\,, \quad
\hat{g}^{W}_{\mcd_2 \mcu_1} =  c_\mcu s_\mcd\,, \quad
\hat{g}^{W}_{\mcd_2 \mcu_2} =  s_\mcu s_\mcd\,,
\end{eqnarray}
where $
g_V^f =  T^3_f - Q_f s_W^2$.

\section{Constraints on the type-I-II 2HDM}

\label{sec:constraints}

Before investigating the allowed parameter space by the current data,
we make some comments on the decays of the VLQs in the type-I-II 2HDM.
As being colored fermions,
the VLQs are copiously produced through the gluon fusion.
The decays depend on the mixing with the SM fermions.
Here we assume that the mixing is very suppressed, below $\lsim \mco(10^{-7})$,
so that the mixing effects on
the $H^\pm \to W^\pm \gm$ and FCNC processes are negligible.
Then the decays of the VLQs are determined by the Yukawa interactions with the Higgs fields.
For example, the $(X,T)$ case in the type-I-II model has
\bea
\label{eq:XT:Yukawa}
-\lg = \dt Y_{4u}\;
\bmcq \Phi_2 u_R +
\dt Y_{4d} \; \bar{Q}_L \Phi_1 \mcd + {\rm h.c.}
\eea
Since the first term 
yields the mixing between $T$ and the SM up-type quarks
and the second term
generates the vertices of $X$-$u_i$-$H^\pm$ and $T$-$d_i$-$H^\pm$,
the decays of $X$ and $T$ are $ X\to W^+ u_i/H^+ u_i$ and $T \to Z u_i/h u_i/W^+ d_i/H^+ d_i$.

Another comment is on the most sensitive FCNC process, $b \to s\gm$.
The comparison between the Belle result\,\cite{Saito:2014das} and
the SM calculation with NNLO QCD correction~\cite{Chetyrkin:1996vx,Buras:1997bk,Buras:2002tp,Misiak:2004ew,Neubert:2004dd,Melnikov:2005bx,
Misiak:2006zs,Misiak:2006ab,Czakon:2006ss,Boughezal:2007ny,Ewerth:2008nv,Misiak:2010sk,
Ferroglia:2010xe,Misiak:2010tk,Czakon:2015exa}
generally puts significant constraints on $\mch$ in
the 2HDM~\cite{Ciafaloni:1997un,Ciuchini:1997xe,Borzumati:1998tg,Bobeth:1999ww,Gambino:2001ew,Misiak:2015xwa}.
Since the SM fermions are assigned in type-I and the VLF contributions are assumed negligible, 
the process $b \to s \gm$ does not practically constrain $\mch$ for $\tb>2$~\cite{Misiak:2017bgg}.

Now we study other constraints on the model such as the Higgs precision data,
the direct searches for the charged Higgs boson and VLFs at the LHC,
and
the electroweak oblique parameters.
Based on the results, we shall suggest a benchmark scenario for this model.

\subsection{Constraints from the LHC Higgs precision data}

The new VLFs change the loop-induced $h$-$g$-$g$ and $h$-$\gamma$-$\gamma$ vertices
which are stringently constrained by
the current Higgs precision measurement.
New physics effects are usually parametrized by the coupling modifier $\kp_i$.
Since $\kp_\gm$ is mainly from $W^\pm$ loop,
the most sensitive one is $\kp_g$,
which the VLFs change into
\begin{eqnarray}
\kappa_g = 1 + \frac{v}{ A^H_{1/2}(\tau_t)} \sum_{\mcf} \sum_i \frac{y^h_{\mcf_i \mcf_i}}{M_{\mcf_i}} 
\,A^H_{1/2}(\tau_{\mcf_i}),
\end{eqnarray}
where the loop function $A^H_{1/2}(\tau)$ is given in Ref.~\cite{Djouadi:2005gi},
$\tau_f = m_h^2/m_f^2$, $\mcf=\mcu,\mcd$, and $i=1,2$.
As explicitly shown in Eq.~(\ref{eq:Yukawa:coupling:m:eigen}),
the vectorlike nature of new fermions yields
\bea
y^h_{\mcf_1 \mcf_1} = -y^h_{\mcf_2 \mcf_2}.
\eea
Unless $M_{\mcf_1}$ is very different from $M_{\mcf_2}$,
the contribution from $\mcf_1$ is considerably canceled by that from $\mcf_2$.
The ATLAS and CMS combined result at $2\sg$~\cite{Khachatryan:2016vau},
$0.6 < |\kappa_g| <1.12$, is satisfied in most of the parameter space.

\subsection{Constraints from direct searches at the LHC}

The VLQ searches have been performed by both
ATLAS~\cite{Aad:2015mba,Aad:2015kqa,Aad:2015voa,Aad:2016qpo,Aad:2016shx,Aaboud:2017qpr,Aaboud:2017zfn,
Aaboud:2018xuw,Aaboud:2018uek,Aaboud:2018saj,Aaboud:2018wxv,Aaboud:2018pii} and
CMS~\cite{Chatrchyan:2013wfa,Khachatryan:2015gza,Khachatryan:2015oba,Khachatryan:2016vph,
Sirunyan:2016ipo,Sirunyan:2017ezy,Sirunyan:2017tfc,Sirunyan:2017usq,Sirunyan:2017ynj,Spiezia:2017ueo}
Collaborations.
No signal of any VLQ gives the lower bound on the VLQ mass,
depending on the assumption of the decay modes.
If $T$ ($B$) decays only into $Zt/Wb/ht$  ($Zb/Wt/hb$),
the bound is very stringent like
$M_T > 1.31\tev$ ($M_B > 1.03\tev$)~\cite{Aaboud:2018pii}.
The mass bounds are relaxed by allowing other decay channels of the VLQs~\cite{Chala:2017xgc}.
For example, if $T$ or $B$ decays into a light quark $q$ associated with $W^\pm$ and $Z$,
the mass bound is 
$M_\mcq > 690\gev$~\cite{Aad:2015tba}.
If $H^\pm q$ mode is additionally open,
the VLQ mass bound can be lower.
As for the VLL, multi-leptonic event searches at the LHC lead to
$M_L \gtrsim 300\gev$ from the ATLAS data~\cite{Dermisek:2014qca}
and $M_L \gtrsim 270\gev$ from the CMS data~\cite{Falkowski:2013jya}.
For the numerical analysis, therefore,
we consider two cases of $M_Q = 600\gev$ and $M_Q = 1.3\tev$ for the VLQs,
and one case of $M_L=300\gev$ for the VLLs.
We shall also consider the LHC direct search bound on the charged Higgs boson,
$\br(t \to b H^+)\times \br(H^+ \to \tau^+\nu)$~\cite{Aaboud:2018gjj}.

\subsection{Constraints from the electroweak oblique parameter $\hat T$}
The electroweak precision test puts
one of the strongest indirect-constraints on new fermions that affect 
the Peskin-Takeuchi oblique parameters $S$, $T$, and $U$~\cite{Peskin:1991sw}.
For more general parametrization,
Barbieri \textit{et al.} extended the parameters into $\hat S$, $\hat T$, $W$, and $Y$~\cite{Barbieri:2004qk},
which are defined as follows.
We begin with $\Pi_{VV'}^{\mu\nu}(q^2)$, the transverse vacuum polarization amplitude of the gauge boson.
Expanding the $g^{\mu\nu}$ term of $\Pi_{VV'}^{\mu\nu}(q^2)$  up to quadratic order as
\bea
\Pi_{VV'}(q^2) \simeq \Pi_{VV'}(0) + q^2 \Pi'_{VV'}(0) + \frac{(q^2)^2}{2} \Pi^{\prime \prime}_{VV'}(0) + \cdots,
\eea
we define $\hat S$, $\hat T$, $W$, and $Y$ as
\begin{eqnarray}
\label{eq:def:STU}
\hat S &=&
\frac{g}{g'}  \Pi^\prime_{W_3 B}(0)\,,\\[3pt] \nn
\hat T &=& \frac{1}{m_W^2 }\left[ \Pi_{W_3 W_3}(0) - \Pi_{W^+ W^-}(0)\right] \,, \\[3pt] \nn
Y &=& \frac{m_W^2 }{2}  \Pi^{\prime \prime}_{BB}(0) \,, \\[3pt] \nn
 W &=& \frac{m_W^2 }{2} \Pi^{\prime \prime}_{W_3W_3}(0)\,.
\end{eqnarray}
The traditional Peskin-Takeuchi parameters $S$ and $T$ are related with $\hat{S}$ and $\hat{T}$ as
\bea
S = \frac{4 s_W^2}{\al} \hat{S},
\quad
T =\frac{\hat{T}}{\al}.
\eea
The current experimental constraints are~\cite{Barbieri:2004qk,Tanabashi:2018oca}
\begin{eqnarray}
\hat S &=& (0.0\pm 1.3)\times 10^{-3}\,,\nn \\
\hat T &=& (5.4\pm 9.3)\times 10^{-4}\,, \nn \\
     W &=& (0.1\pm 1.2)\times 10^{-3}\,, \nn \\
     Y &=& (-0.4\pm 0.8)\times 10^{-3}\,.
\end{eqnarray}

We focus on the most sensitive oblique parameter $\hat{T}$ to new fermions.
The discussions on $\hat{S}$, $Y$, and $W$ are in Appendix \ref{appendix:Pi}.
For the general vector and axial-vector gauge couplings
defined by $\lg= V_\mu \bar \psi (g_V \gamma^\mu+g_A \gamma_5 \gamma^\mu) \psi$,
$\Pi_{V V'}(0)$ from a single diagram
mediated by two fermions with masses $m_a$ and $m_b$ is~\cite{Cynolter:2008ea}
\begin{eqnarray}
\Pi(m_a,m_b,0) &=& \frac{1}{4\pi^2} \left[  (g_V^2 +g_A^2) \tilde \Pi_{V+A}(m_a,m_b,0)
+ (g_V^2-g_A^2)\tilde \Pi_{V-A}(m_a,m_b,0)\right] ,
\end{eqnarray}
where the subscript $VV'$ in $\Pi_{VV'}$ is omitted for simplicity and $\tilde{\Pi}_{V\pm A}(m_a,m_b,0)$ are
\begin{eqnarray}
\tilde \Pi_{V+A}(m_a,m_b,0) &=& -\frac{1}{2}(m_a^2+m_b^2) \left[ {\rm Div} + \ln \Big( \frac{\mu^2}{m_a m_b}\Big) \right]
\\ \nn
&& -\frac{1}{4} (m_a^2 +m_b^2) - \frac{(m_a^4+m_b^4)}{4(m_a^2-m_b^2)} \ln \Big(\frac{m_b^2}{m_a^2}\Big)\,, \\
\tilde \Pi_{V-A}(m_a,m_b,0) &=& m_a m_b \left[ {\rm Div} +\ln \Big(\frac{\mu^2}{m_a m_b}\Big)
+1 +\frac{(m_a^2+m_b^2)}{2(m_a^2-m_b^2)} \ln \Big(\frac{m_b^2}{m_a^2}\Big) \right] \,.\nn
\end{eqnarray}
Here ${\rm Div} = 1/\epsilon + \ln 4\pi - \gamma_\epsilon$ is the divergence term in the dimensional regularization,
$\es = (4-D)/2$,
and $\mu$ is the renormalization scale.
The divergences from the VLF contributions are properly canceled out and there is no $\mu$ dependence on $\hat T$.
The vectorlike nature of new fermions makes $\hat{T}$ depend only on $\tilde \Pi_V$, defined by
\bea
\label{eq:Pi:V}
\tilde \Pi_V =  \tilde \Pi_{V+A} + \tilde \Pi_{V-A}.
\eea
Then, $ \hat T$ in our model is
 \begin{eqnarray}
 \label{eq:That}
 \hat T = \frac{g^2 N_C}{16\pi^2 m_W^2} && \bigg[ ~
2s_\mcu^2 c_\mcu^2 \tilde \Pi_V(M_{\mcu_1},M_{\mcu_2},0)
             + 2s_\mcd^2 c_\mcd^2 \tilde \Pi_V(M_{\mcd_1},M_{\mcd_2},0) \nn \\[3pt]
 &&    - \, 2c_\mcu^2 c_\mcd^2 \tilde \Pi_V(M_{\mcu_1},M_{\mcd_1},0)
             - 2s_\mcu^2 s_\mcd^2 \tilde \Pi_V(M_{\mcu_2},M_{\mcd_2},0) \nn \\[3pt]
 &&   - \, 2c_\mcu^2 s_\mcd^2 \tilde \Pi_V(M_{\mcu_1},M_{\mcd_2},0)
             - 2s_\mcu^2 c_\mcd^2 \tilde \Pi_V(M_{\mcu_2},M_{\mcd_1},0)
 \bigg]  ,
 \end{eqnarray}
where $N_C=3~(1)$ for the VLQ (VLL).

\begin{figure}[h] \centering
\includegraphics[width=0.5\textwidth]{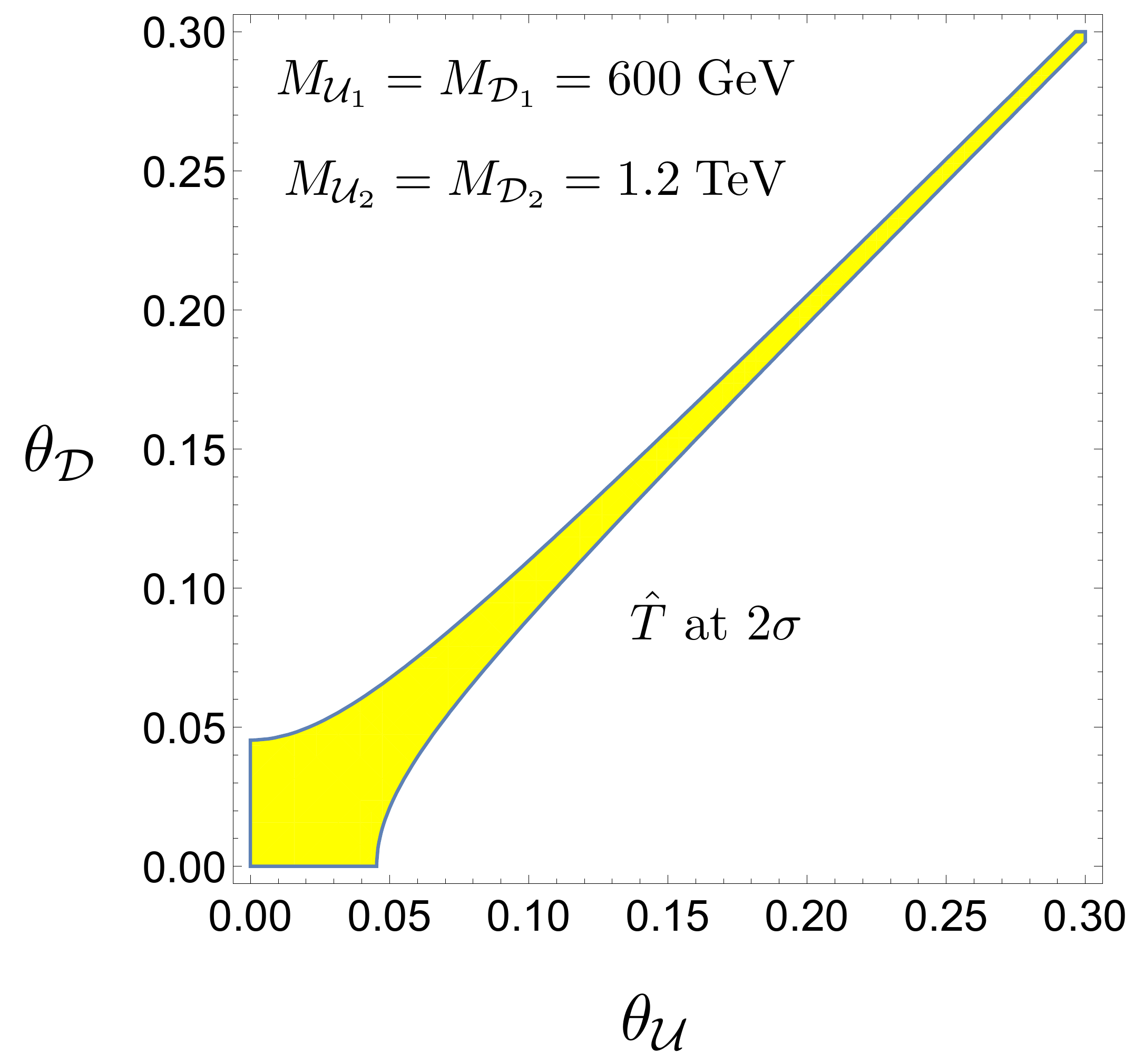}
\caption{
The allowed region of $(\theta_\mcu, \theta_\mcd)$ at $2\sg$
by the electroweak oblique parameter $\hat T$.
We set $M_{\mcu_1}=M_{\mcd_1}=600\gev$ and $M_{\mcu_2}=M_{\mcd_2}=1.2\tev$.
}
\label{fig-That-Allowed}
\end{figure}

It is generally known that the small $\hat{T}$ prefers very degenerate masses of the new fermions in the loop,
which is clearly seen from
\bea
\label{eq:ma=mb}
\lim_{m_a \to m_b} \tilde{\Pi}_{V+A}(0)=-m_a^2
\left[
{\rm Div} + \ln \lf \frac{\mu^2}{m_a^2} \ri
\right] = - \lim_{m_a \to m_b} \tilde{\Pi}_{V-A}(0).
\eea
As will be shown, however, the crucial condition for the enhancement of $\br(H^\pm \to W^\pm \gm)$
is the sizable mass difference between the up-type and down-type VLFs.
It seems that the $\hat{T}$ constraint excludes the possibility of the enhancement.
Here comes the advantage of our model with \emph{vectorlike} $SU(2)_L$ doublet \emph{and} singlet fermions.
The new fermion spectrum includes $\mcu_1$, $\mcu_2$, $\mcd_1$, and $\mcd_2$,
leading to six terms in Eq.~(\ref{eq:That}).
Now each term can be sizable while
$\hat{T}$ is kept small if the first two terms
are canceled by the last four terms.
We find that this cancellation occurs when $M_{\mcu_i} \approx M_{\mcd_i}$
and $\theta_\mcu \approx \theta_\mcd$.
In Fig.~\ref{fig-That-Allowed},
we show the $2\sg$ allowed region
of $(\theta_\mcu, \theta_\mcd)$
by the electroweak oblique parameter $\hat T$ for
$M_{\mcu_1}=M_{\mcd_1}=600\gev$ and $M_{\mcu_2}=M_{\mcd_2}=1.2\tev$.
In conclusion, we find the following simple ansatz to satisfy $\hat{T}=0$:
\bea
\label{eq:ansatz}
M_{\mcu_1}=M_{\mcd_1}, \quad M_{\mcu_2}=M_{\mcd_2}, \quad \theta_\mcu =\theta_\mcd.
\eea

\subsection{Benchmark scenario for the numerical analysis}
Considering all of the constraints discussed above,
we take the following benchmark scenario:
\begin{eqnarray}
\label{eq:benchmark}
&& \sba = 1, \quad \hbox{(alignment limit)},
\\[5pt] \nn
&&M_{\mcu_1}=M_{\mcd_1} =
\left\{
\begin{array}{ll}
600\gev ~{\rm or}~ 1.3\tev, & \hbox{for the VLQs}; \\
300\gev, & \hbox{for the VLLs},
\end{array}
\right.
\\[5pt] \nn
&& (Q_\mcu, Q_\mcd)=
\left\{
\begin{array}{ll}
\hbox{VLQ: } &
\left[
	\begin{array}{ll}
	(X,T) &:  (5/3,2/3 ); \\
	(T,B) &: ( 2/3, -1/3 ); \\
	(B,Y)&: (-1/3,-4/3);
	\end{array}
\right.
\\[12pt]
\hbox{VLL: } & ~~(N,E): (0,-1) ,
\end{array}
\right.
\\[5pt]
&& \Delta M \equiv M_{\mcu_2} - M_{\mcu_1} = M_{\mcd_2} - M_{\mcd_1} \subset [ 0 , 1.5 ]\tev, \nn \\[5pt]
\nn
&& \theta_\mcu = \theta_\mcd = 0.2\,,
\end{eqnarray}
where $Q_\mcf$ is the electric charges of the particle $\mcf$.
Note that the ansatz in Eq.~(\ref{eq:ansatz}) relates the up-type Yukawa coupling $Y_\mcu$
with the down-type Yukawa coupling $Y_\mcd$ as
\bea
\label{eq:Yd:Yu}
Y_\mcd = Y_\mcu \,\tb,
\eea
which can be clearly seen from Eq.~(\ref{eq:mixingangle}).

\section{Loop induced decays of the charged Higgs boson}
\label{sec:calculation}

\begin{figure}[h] \centering
\includegraphics[width=0.9\textwidth]{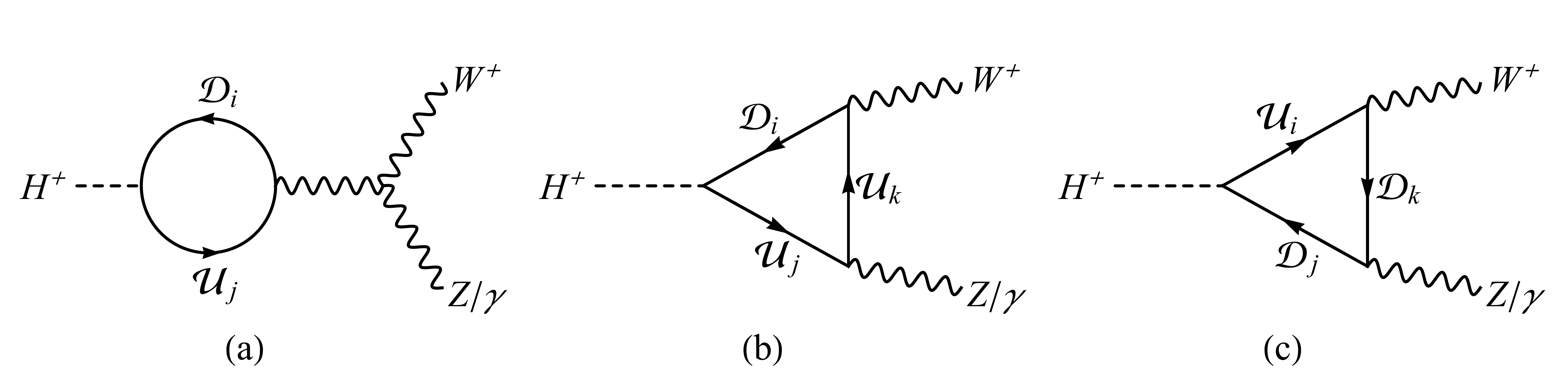}
\caption{
Feynman diagrams for $H^+ \to W^+ \gamma/ W^+ Z$.
Here $\mcu_{i}$ and $\mcd_{i}$ denote the up-type and down-type VLFs as well as the SM $t$ and $b$ quarks,
respectively.
 \label{fig:diagrams}
 }
\end{figure}

In our model,
the decays of $H^\pm \to W^\pm \gamma$ and $H^\pm \to W^\pm Z$ occur
radiatively
through the VLFs as well as the SM top and bottom quarks,
as shown in Fig.~\ref{fig:diagrams}.
The loop-induced decay amplitude of $H^+ \to W^+ V$ ($V =\gamma,Z$) is parametrized by
\begin{equation}
{\cal M} = \frac{    g^2 N_C \mch}{(16\pi^2)\sqrt{2} c_W}
{\cal M}_{\mu \nu} \, \varepsilon^{\mu *}_W \varepsilon^{\nu *}_V ,
\end{equation}
where $N_C$ is the color factor of the fermion in the loop.
We further express ${\cal M}_{\mu\nu}$ in terms of three dimensionless form-factors ${\cal M}_{1,2,3}$ as
\begin{equation}
\label{eq:FF}
{\cal M}_{\mu \nu} = g_{\mu\nu} {\cal M}_1 + \frac{p_{2 \mu} p_{1 \nu}}{\mch^2} {\cal M}_2
+ i \epsilon_{\mu\nu\rho\sigma} \frac{p_{2}^{\rho} p_{1}^{\sigma}}{\mch^2} {\cal M}_3\,,
\end{equation}
where $p_1$ and $p_2$ are the momenta of $W^\pm$ and $V$ respectively.

In our model, each ${\cal M}_q$ ($q=1,2,3$)
receives the contributions
from various VLF combinations through the Feynman diagrams (a), (b), and (c) in Fig.~\ref{fig:diagrams}.
Since there are
two up-type VLFs and two down-type VLFs ($\mcu_{1,2}$ and $\mcd_{1,2}$),
we index the form-factors by the superscripts for the diagrams
and by the subscripts for the VLFs:
\begin{equation}
\label{eq:FFsum}
{\cal M}_q = \sum_{i,j}{\cal M}_{q,\,ij}^{\rm (a)}
+\sum_{i,j,k}\left[ {\cal M}_{q,\,ijk}^{\rm (b)}+{\cal M}_{q,\,ijk}^{\rm (c)}\right] \,,~~ \hbox{for } q=1,2,3\,.
\end{equation}
We summarize the indices of $i$, $j$, and $k$
for $W^\pm\gm$ and $W^\pm Z$
in Table \ref{tab:index}.

\begin{table}[h!]
\begin{center}
  {\renewcommand{\arraystretch}{1.2}
\begin{tabular}{|l|c|c|}
\hline
& $H^+ \to W^+ \gm$ &~~ $H^+ \to W^+ Z$ ~~~\\
\hline
~~ diagram (a): $(i,j)$~~ & $(11), (12), (21), (22)$   ~~&~~ $(11), (12), (21), (22)$ ~~~\\
\hline
~~ diagram (b) and (c): ~~& \multirow{2}{*}{$(111), (122), (211), (222)$}   ~~&~~  (111), (112), (121), (122)~~\\
~~~~~ $(i,j,k)$ & &~~  (211), (212), (221), (222)~~\\
\hline
\end{tabular}
}
\caption{\label{tab:index}
 \baselineskip 3.5ex
 The values of indices of VLQs for each diagram.}
\end{center}
\end{table}

For $W^+ \gamma$ decay, the Ward-identity of
$p_{2}^{\,\nu} {\cal M}_{\mu\nu}=0$ from the gauge invariance
relates ${\cal M}_1$ with ${\cal M}_2$ as
\begin{equation}
\label{eq:WardIdentity}
{\cal M}_1 = - \frac{1}{2} \Big( 1 -\mu_W\Big) {\cal M}_2,~~~
{\rm for}~H^+ \to W^+ \gamma\,,
\end{equation}
where $\mu_i = m_i^2/\mch^2$.
The partial decay rate for $H^+ \to W^+ \gamma$ is
\begin{equation}
\label{eq:dwWgamma}
\Gamma(H^+ \to W^+ \gamma) = \frac{g^4 N_C^2 }{2^{14} \pi^5 c_W^2}
\, \mch
\lf 1 - \mu_W\ri^3
\left[  \left| {\cal M}_2 \right|^2 +\left| {\cal M}_3 \right|^2 \right]\,.
\end{equation}
The partial decay rate for $H^+ \to W^+ Z$ is \begin{eqnarray}
\label{eq:GammaHp2WZ}
\Gamma(H^+ \to W^+ Z) &=& \frac{g^4 N_C^2 \beta }{2^{14} \pi^5 c_W^2}\, \mch
\Bigg[ \lf 6 + \frac{\beta^2}{2 \mu_W \mu_Z}\ri \left| {\cal M}_1 \right|^2
+\frac{\beta^4 }{8\mu_W \mu_Z}\left| {\cal M}_2 \right|^2
+\beta^2\left| {\cal M}_3 \right|^2
\nn \\
&&~~ + \frac{\beta^2}{2}\Big(\frac{1}{\mu_W \mu_Z}
 - \frac{1}{\mu_W} - \frac{1}{\mu_Z} \Big) {\rm Re}\big( {\cal M}_1 {\cal M}_2^* \big)
\Bigg]\,,
\end{eqnarray}
where
$
\beta = \sqrt{\lf 1-\mu_W -\mu_Z \ri^2 - 4\mu_W \mu_Z }
$ and the Ward identity in Eq.~(\ref{eq:WardIdentity}) does not apply.
Note that $\Gm(H^\pm \to W^\pm Z)$ increases
with $\mch$ because of the longitudinal polarization contribution which is proportional to
$1/(\mu_W \mu_Z)$, i.e.,
$\mch^4/m_W^2 m_Z^2$.
The detailed expressions of ${\cal M}_1$, ${\cal M}_2$, and ${\cal M}_3$
from the VLF loops as well as the SM $t$ and $b$ quark loops
are shown in Appendix \ref{appendix:form:factors}.
Our calculation of the VLF contributions
is new.
We checked that our expressions for the SM contributions
are numerically consistent with those in Ref.~\cite{CapdequiPeyranere:1990qk}.

\begin{figure}[h] \centering
\includegraphics[width=0.65\textwidth]{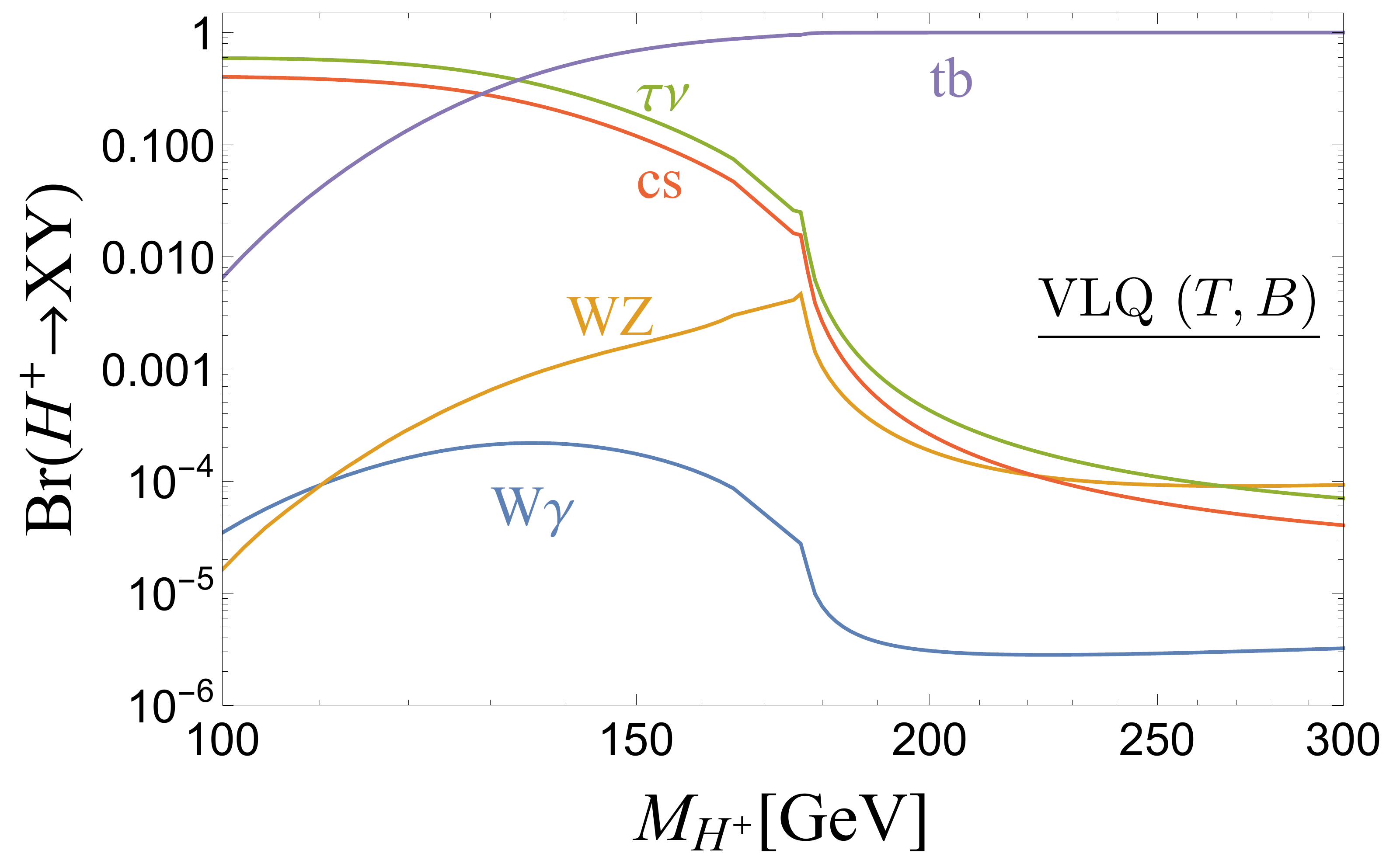}
\caption{
\baselineskip 3.5ex
The branching ratios of the charged Higgs boson
in the type-I-II 2HDM  as a function of
$\mch$. 
As a representative, we take the VLQ $(T,B)$ case.
We set
$M_{\mcu_1} = M_{\mcd_1}=600 \gev$, $\Dt M = 600\gev$, $\tb=10$, and
$\theta_\mcu=\theta_\mcd=0.2$. 
}
\label{fig-HpBR2HDMVLQTB}
\end{figure}

In Fig.~\ref{fig-HpBR2HDMVLQTB},
we show the branching ratios of $H^\pm$ as a function of
$\mch$ for $M_{\mcu_1} = M_{\mcd_1}=600 \gev$, $\Dt M = 600\gev$, $\tb=10$,
and $\theta_\mcu=\theta_\mcd=0.2$.  
Here we take the VLQ $(T,B)$ case as a representative.
The main decay modes of the charged Higgs boson in the 2HDM are still fermionic,
$t^{(*)} b$ for $\mch \gsim 135\gev$ and $\tau\nu$ for lighter $\mch$.
Nevertheless the radiative decays of $W\gm$ and $WZ$ modes
are not negligible for the intermediate-mass charged Higgs boson:
$\br(H^\pm \to W^\pm \gm)$ reaches the maximum at $\mch \simeq 135\gev$
and $\br(H^\pm \to W^\pm Z)$ becomes the largest at $\mch \simeq m_t$.

\begin{figure}[h] \centering
\includegraphics[width=0.45\textwidth]{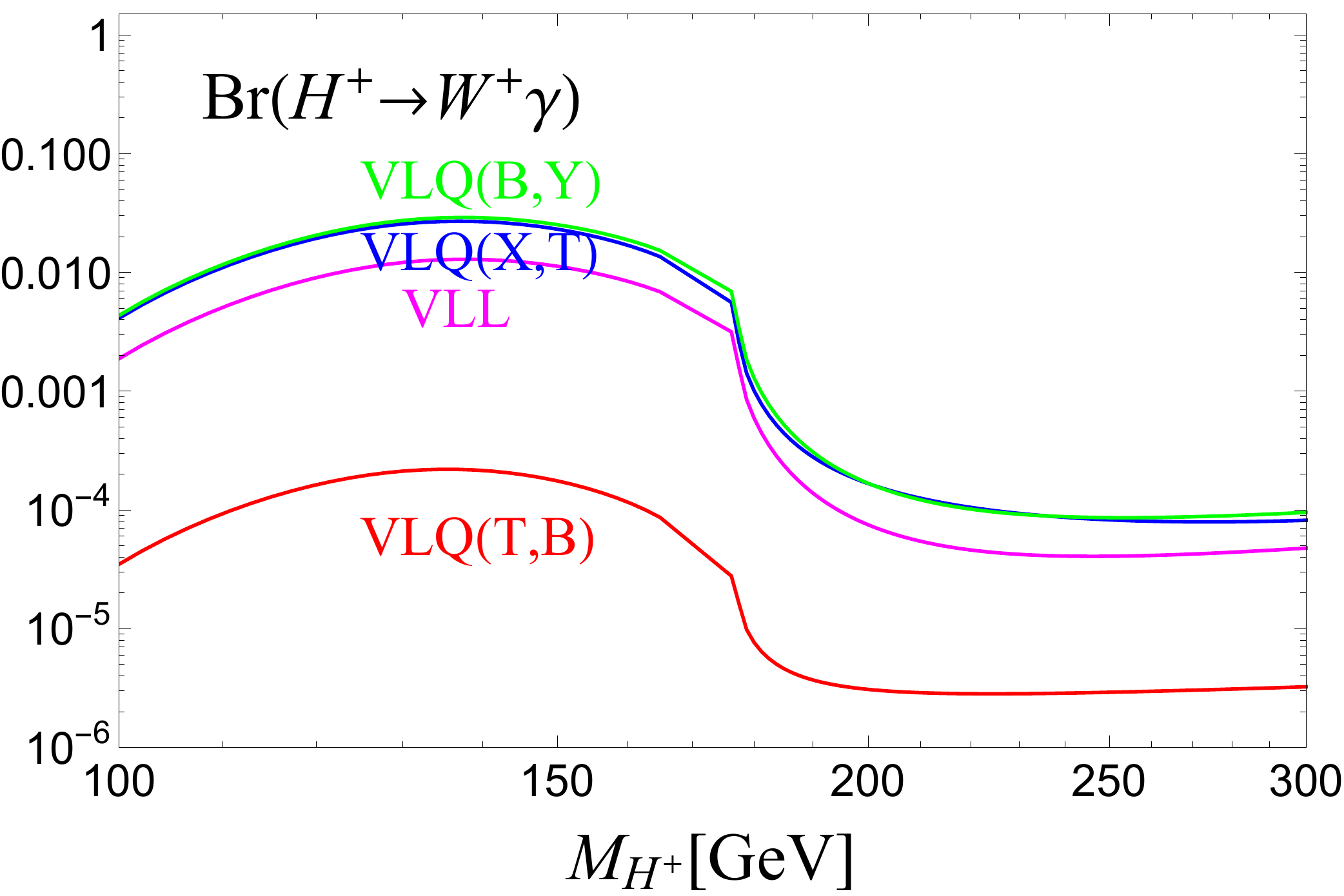}
\includegraphics[width=0.45\textwidth]{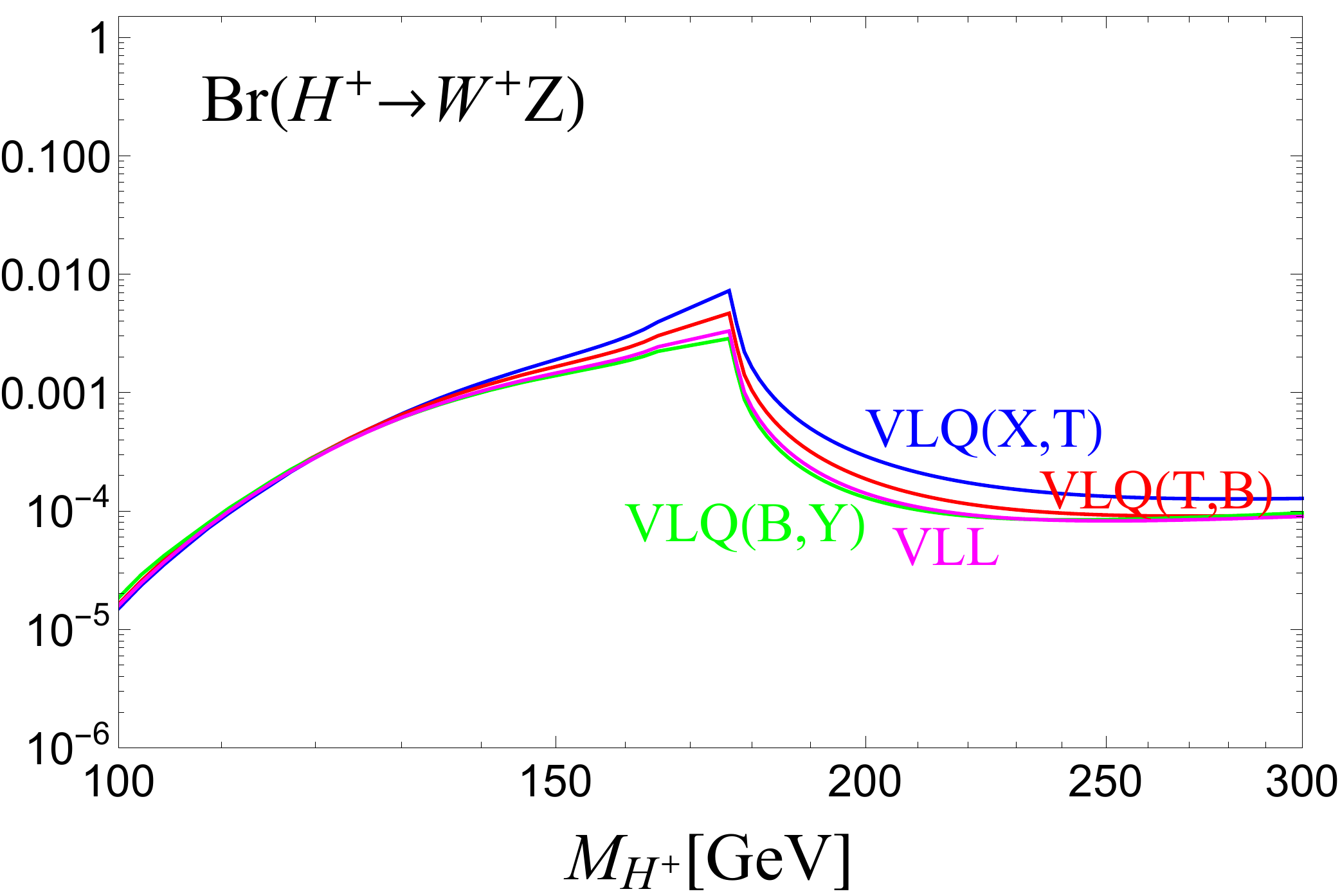}
\caption{
\baselineskip 3.5ex
$\br(H^\pm \to W^\pm \gm)$ (left panel)
and
$\br(H^\pm \to W^\pm Z)$ (right panel) as a function of $\mch$
for the VLQ $(X,T)$, $(T,B)$, $(B,Y)$, and VLL $(N,E)$ cases.
We set
$M_{\mcu_1} = M_{\mcd_1}=600 \gev$, $\Dt M = 600\gev$, $\tb=10$, and
$\theta_\mcu=\theta_\mcd=0.2$. 
}
\label{fig-HpBRWgammaVLQ}
\end{figure}

Fig. \ref{fig-HpBRWgammaVLQ} shows $\br(H^\pm \to W^\pm \gm)$ (left panel)
and $\br(H^\pm \to W^\pm Z)$ (right panel)
for other VLF cases.
Both $\br(H^\pm \to W^\pm\gm)$ 
and $\br(H^\pm \to W^\pm Z)$ are suppressed above the $tb$ threshold
because the $tb$ decay mode is very dominant.
For the intermediate-mass charged Higgs boson ($140\gev \lsim \mch \lsim 180\gev$), 
the radiative decays are sizable.
In the details,
the $W\gm$ and $WZ$ modes are different.
$\br(H^\pm \to W^\pm Z)$ is very similar for
all of the VLF cases,
which reaches its maximum of the order of $\mco(10^{-3})$ at the $tb$ threshold.
On the other hand, 
the VLF contributions to $\br(H^\pm \to W^\pm\gm)$ vary dramatically according to the quantum numbers of the VLFs,
although its shape 
as a function of $\mch$ is similar for all of the VLF cases.
The VLQ $(X,T)$, $(B,Y)$, and VLL
have very large $\br(H^\pm \to W^\pm\gm)$,
exceeding $\mco(10^{-2})$ at $\mch \simeq 135\gev$.

\begin{figure}[h] \centering
\includegraphics[width=0.49\textwidth]{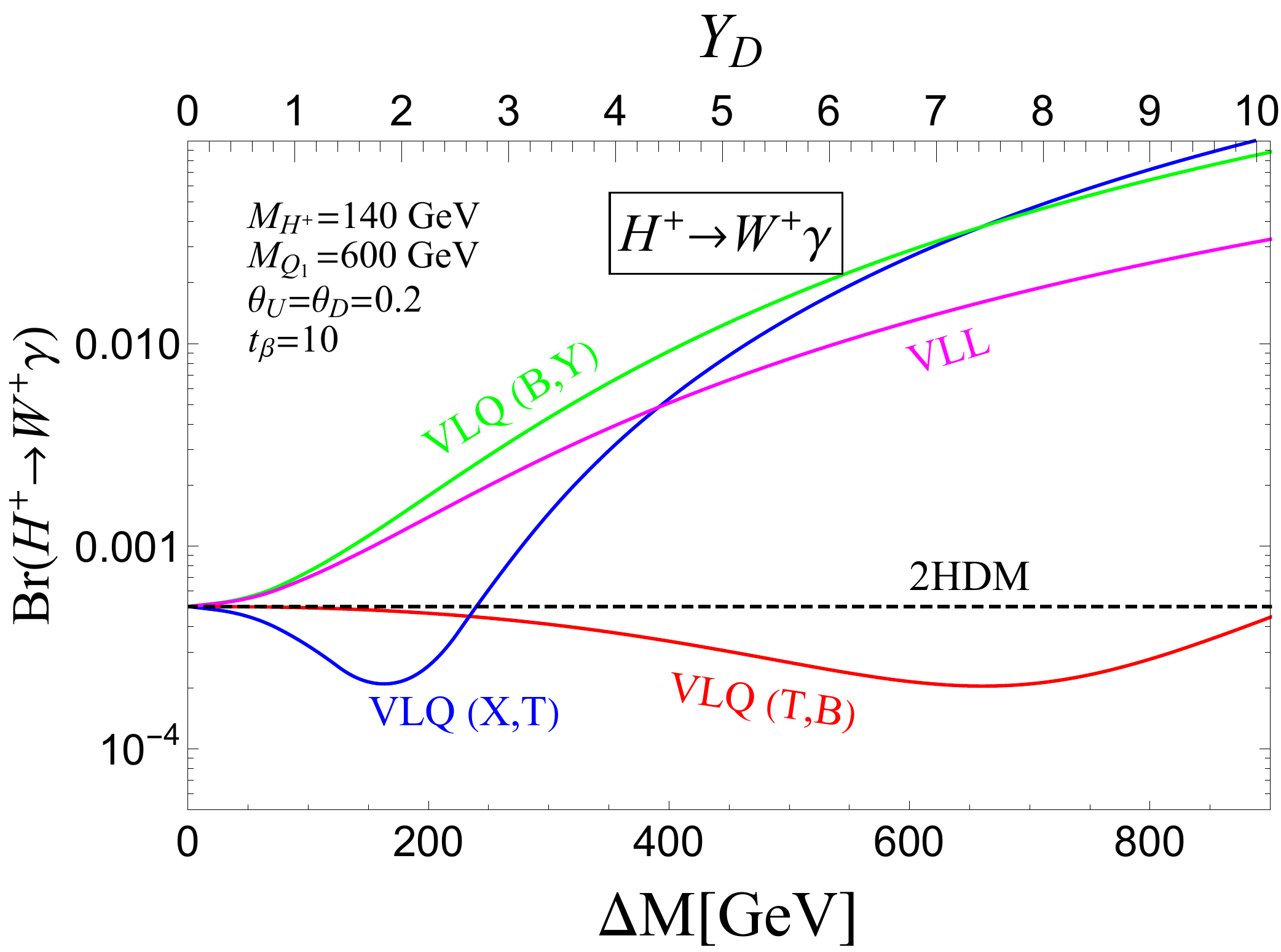}
\includegraphics[width=0.49\textwidth]{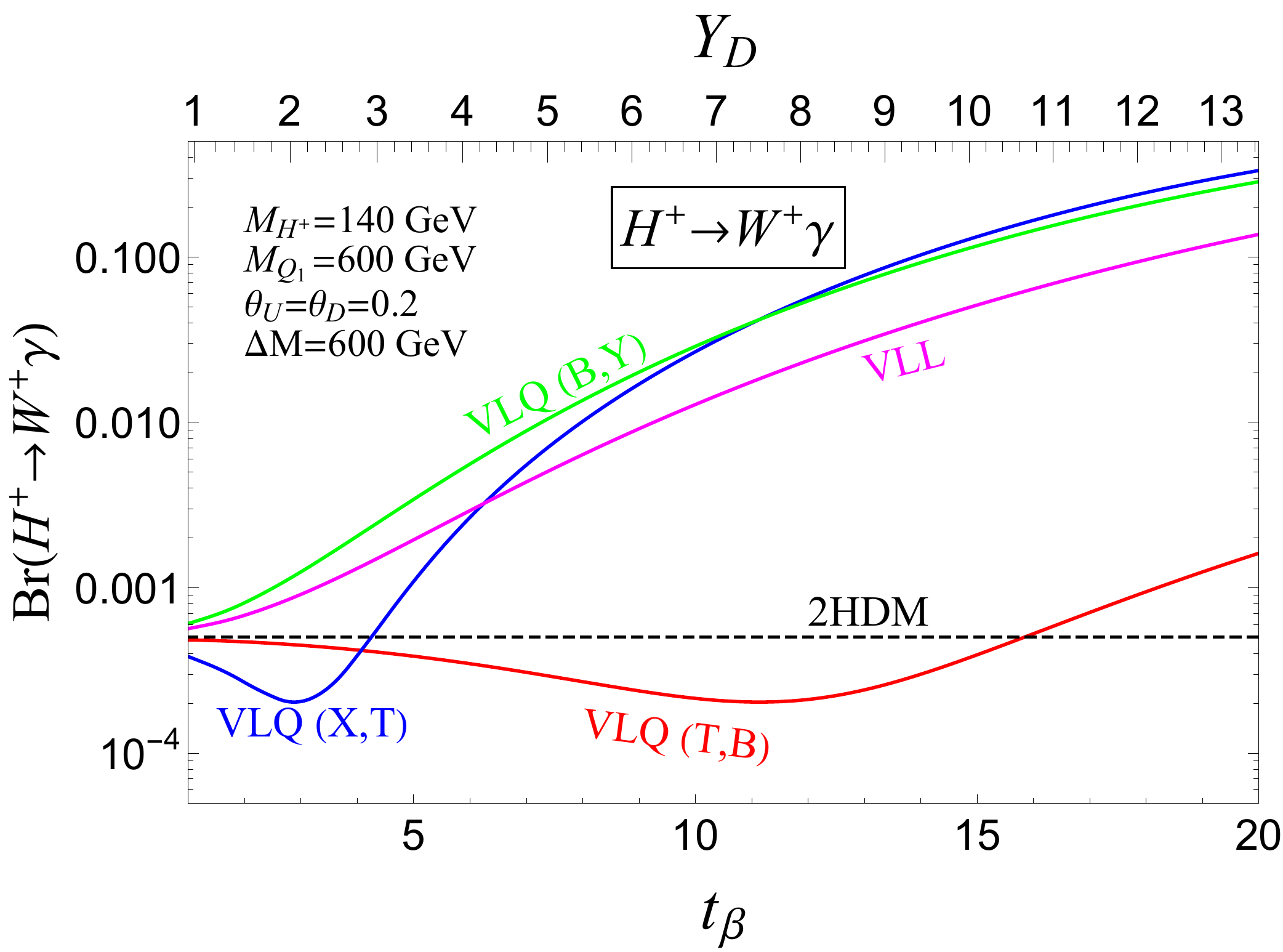}
\caption{
\baselineskip 3.5ex
$\br(H^+\to W\gamma)$ as a function of $\Delta M (\equiv M_{\mcf_2}-M_{\mcf_1})$
for the fixed $\tb=10$
(left panel) and $\tb$ for the fixed $\Dt M=600\gev$ (right panel).
We set $\mch=140\gev$,
$M_{\mcu_1} = M_{\mcd_1}=600 \gev$ for the VLQs,
$M_{\mcu_1} = M_{\mcd_1}=300 \gev$ for the VLLs,
and $\theta_\mcu=\theta_\mcd=0.2$.
The dashed lines represent the results in the type-I 2HDM without VLFs.
}
\label{fig:BRWa:small:M}
\end{figure}

In order to see the parameter dependence of $\br(H^\pm \to W^\pm\gm)$,
we show the branching ratios of $H^+ \to W^+ \gamma$
as a function of $\Delta M (\equiv M_{\mcf_2}-M_{\mcf_1})$ for the fixed $\tb=10$ (left panel),
and as a function of $\tb$ for the fixed 
$\Dt M=600\gev$
(right panel) in Fig.~\ref{fig:BRWa:small:M}.
We set $\mch=140\gev$, $M_{\mcu_1} = M_{\mcd_1}=600 \gev$ for the VLQs,
$M_{\mcu_1} = M_{\mcd_1}=300 \gev$ for the VLLs,
 and $\theta_\mcu = \theta_\mcd = 0.2$.
Since  $Y_\mcd(=\tb Y_\mcu)$ is determined by $\Dt M$, $\tb$, $\theta_{\mcu}$ and $\theta_{\mcd}$ in our ansatz,
we additionally show the values of $Y_\mcd$ in the plot.
Note that too large $\Dt M$ or $\tb$
endangers the perturbativity of the VLF Yukawa couplings
since the value of $Y_\mcd$
increases with $\Dt M$ and $\tb$.

Let us discuss the characteristic features of $\br(H^\pm \to W^\pm\gm)$.
In the ordinary 2HDM without the VLFs (dashed lines),
the branching ratio 
is very suppressed like $\sim \mathcal{O}(10^{-4})$.
It would probably be impossible to discover the charged Higgs boson through the $W \gm$ mode
at the LHC.
When the VLFs come in the loop,
the effects are not only dramatic but also very different according to their electric charges.
The $(T,B)$ VLQ contribution destructively interferes with the SM contributions
in the most of the parameter space of $\Dt M$ and $\tb$, yielding 
smaller $\br(H^+\to W\gamma)$ than that without the VLFs.
The $(X,T)$  
contribution is 
destructive for small $\Dt M$ or small $\tb$
but 
rapidly exceeds the SM contributions
for large $\Dt M$ or large $\tb$.
Both $(B,Y)$ 
and $(N,E)$ contributions are always positive.
The most remarkable result is that $\br(H^\pm \to W^\pm\gm)$ is highly enhanced,
except for the $(T,B)$ case:
one order of magnitude enhancement is easily achieved with moderately large $\Dt M$ and $\tb$.
If we push the parameters further up to the marginal point satisfying the perturbativity of $Y_\mcd$,
$\br(H^\pm \to W^\pm\gm)$ can be as large as $\sim 0.1$.

The whole behavior of $\br(H^\pm \to W^\pm \gm)$, especially its sensitive dependence on the VLF electric charges,
is not easy to understand
since it involves the complicated loop effects
from various combinations of the VLFs as in Fig.~\ref{fig:diagrams} as well as the SM quarks.
Nevertheless, we find the reason when the VLF loop effects are dominant.
Since ${\cal M}_3({\rm VLF})=0$ and
${\cal M}_2^{\rm (a)}=0$
(see Appendix   \ref{appendix:form:factors}),
non-vanishing contributions
are from ${\cal M}_2^{\rm (b)}$ and ${\cal M}_2^{\rm (c)}$.
As can be seen in Eq.~(\ref{eq:FormFactor:VLQ:M2}),
${\cal M}_2^{\rm (b)}$ is proportional to $Q_\mcu$ while
${\cal M}_2^{\rm (c)}$ is proportional to $Q_\mcd$.
Here $Q_\mcu$ ($Q_\mcd$) is the electric charge of the up-type (down-type) fermion.
In the $(T,B)$ case, the sign of $Q_\mcu$ is opposite to that of $Q_\mcd$,
which yields substantial cancellation between ${\cal M}_2^{\rm (b)}$ and ${\cal M}_2^{\rm (c)}$.
And the remaining $(T,B)$ contribution destructively interferes with the SM contribution.
Other cases of $(X,T)$, $(B,Y)$, and $(N,E)$ with the same-sign $Q_\mcu$ and $Q_\mcd$
can have large branching ratios.

\begin{figure}[h] \centering
\includegraphics[width=0.49\textwidth]{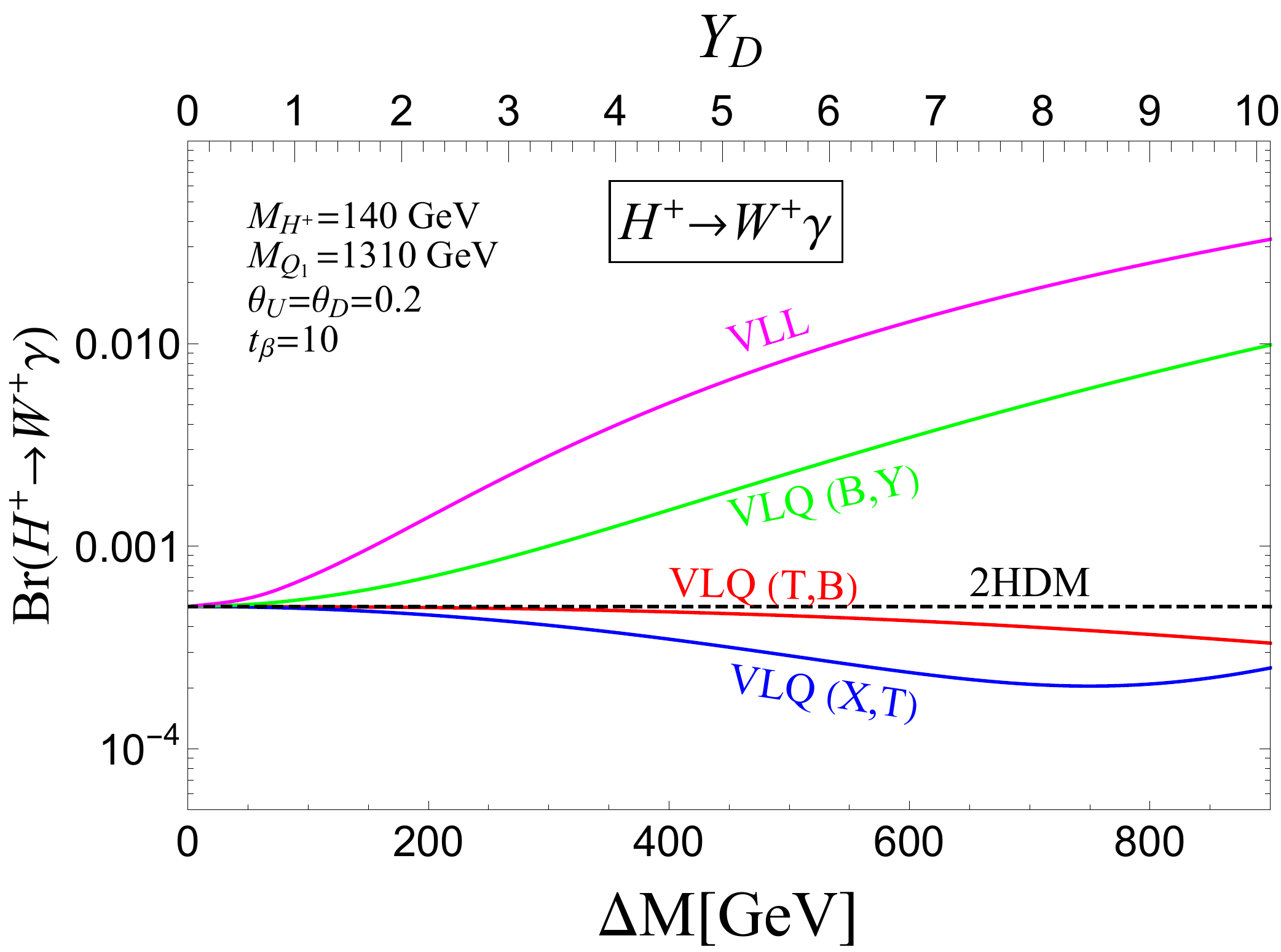}
\includegraphics[width=0.49\textwidth]{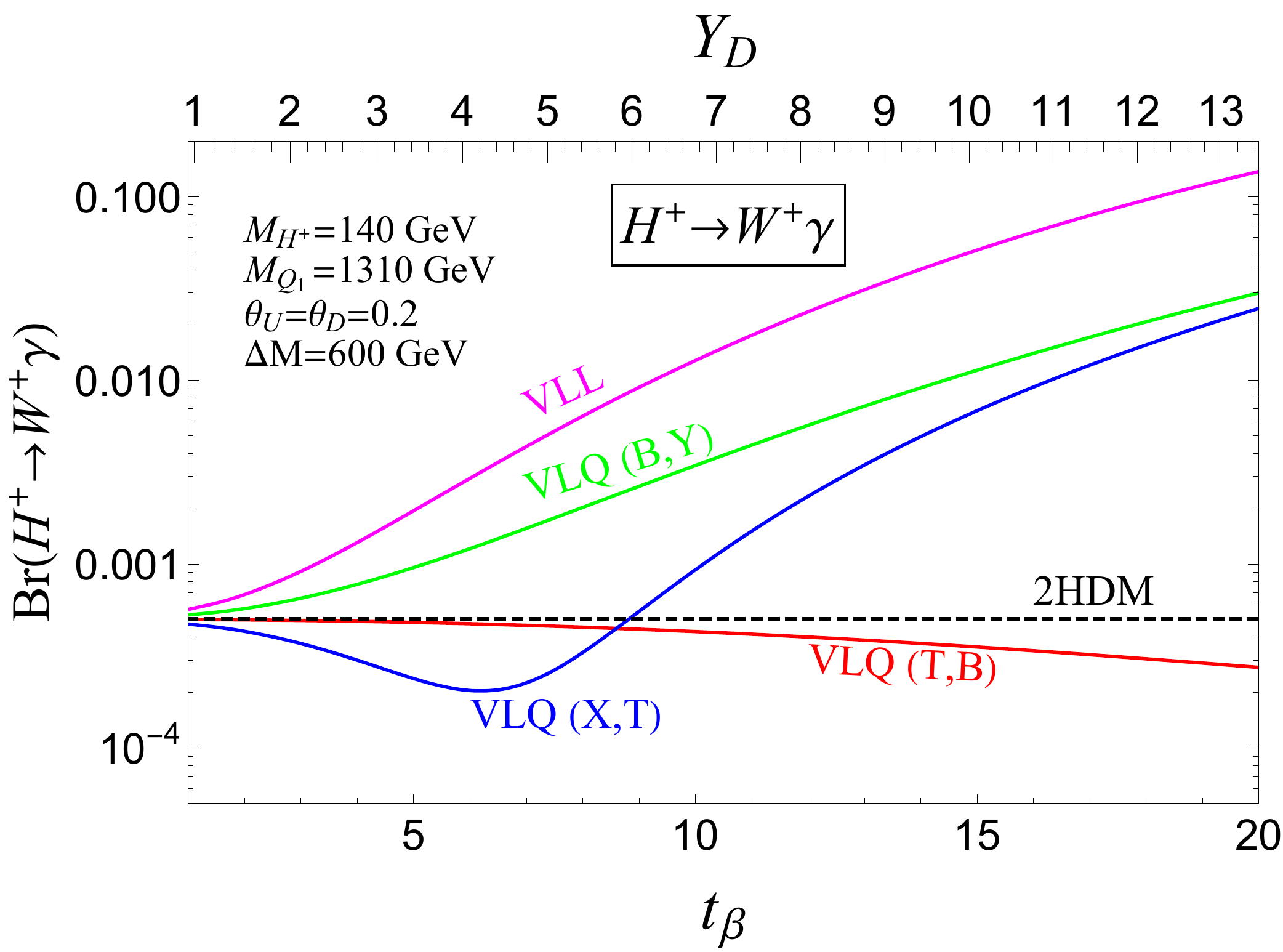}\\
\caption{
\baselineskip 3.5ex
$\br(H^+\to W\gamma)$ with heavy VLFs in the loop as a function of $\Delta M$ 
for the fixed $\tb=10$ (left panels)
and $\tb$ for the fixed $\Dt M=600\gev$ (right panels).
We set $\mch=140\gev$, $M_{\mcu_1}=M_{\mcd_2}=1.31\tev$ for the VLQs,
$M_{\mcu_1} = M_{\mcd_1}=300 \gev$ for the VLLs,
and $\theta_\mcu=\theta_\mcd=0.2$.
The dashed lines represent the result of the type-I 2HDM without VLFs.
}
\label{fig:BRWa:large:M}
\end{figure}

In order to see the VLF mass dependence,
we show $\br(H^\pm \to W^\pm \gamma)$ for heavy VLQs with $M_{\mcu_1}=M_{\mcd_1}=1.31\tev$
in Fig.~\ref{fig:BRWa:large:M}.
The shapes of $\br(H^\pm \to W^\pm \gamma)$ as a function of $\Dt M$ and $\tb$
remain similar to those for light VLQs.
However, the magnitude of $\br(H^+\to W\gamma)$
is reduced significantly, by an order of magnitude, 
for 
heavier VLQs with about twice mass.
But still $\br(H^+\to W\gamma)$ can be an order of magnitude larger than that without the VLFs.

\begin{figure}[h] \centering
\includegraphics[width=0.49\textwidth]{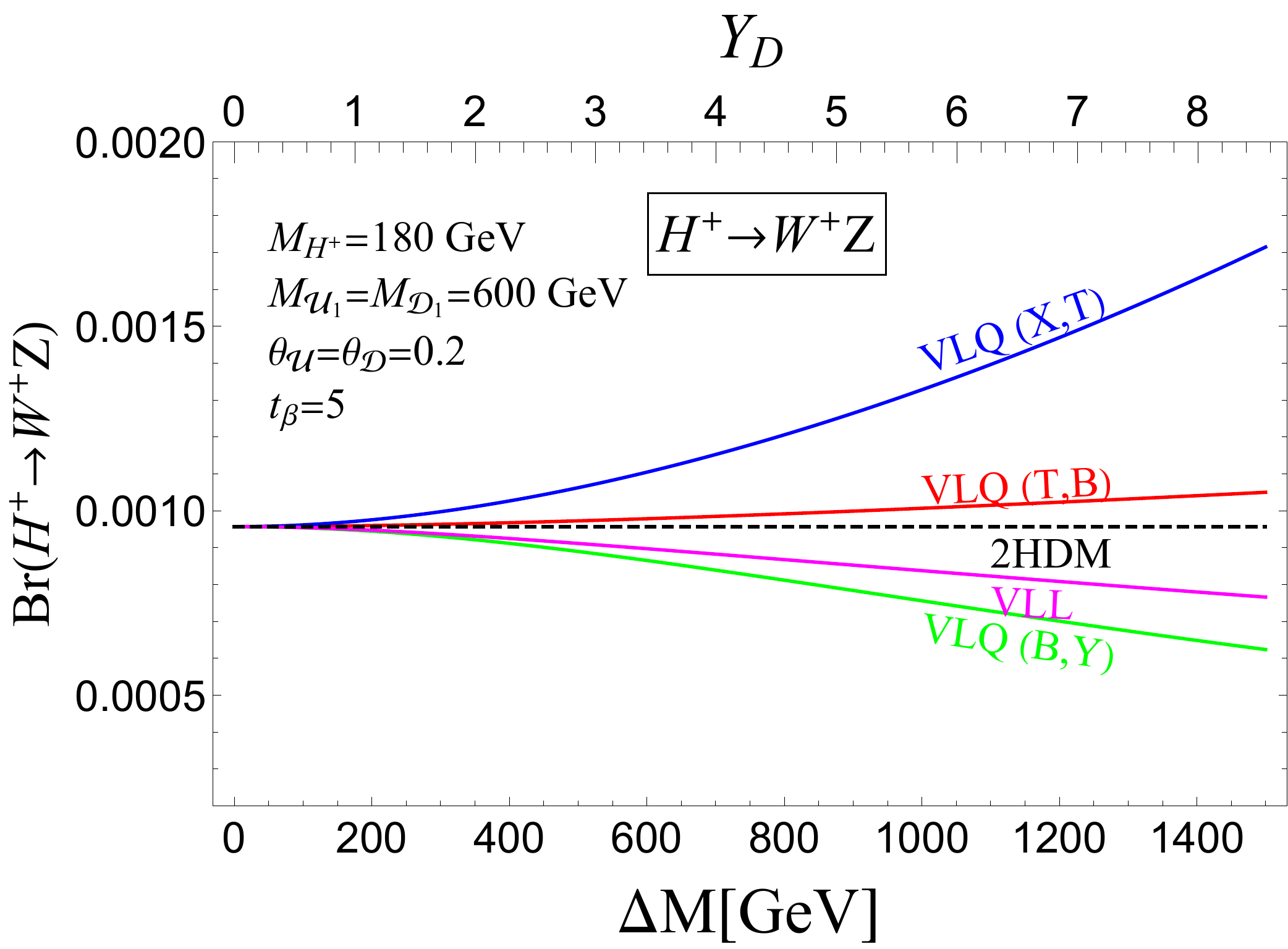}
\includegraphics[width=0.49\textwidth]{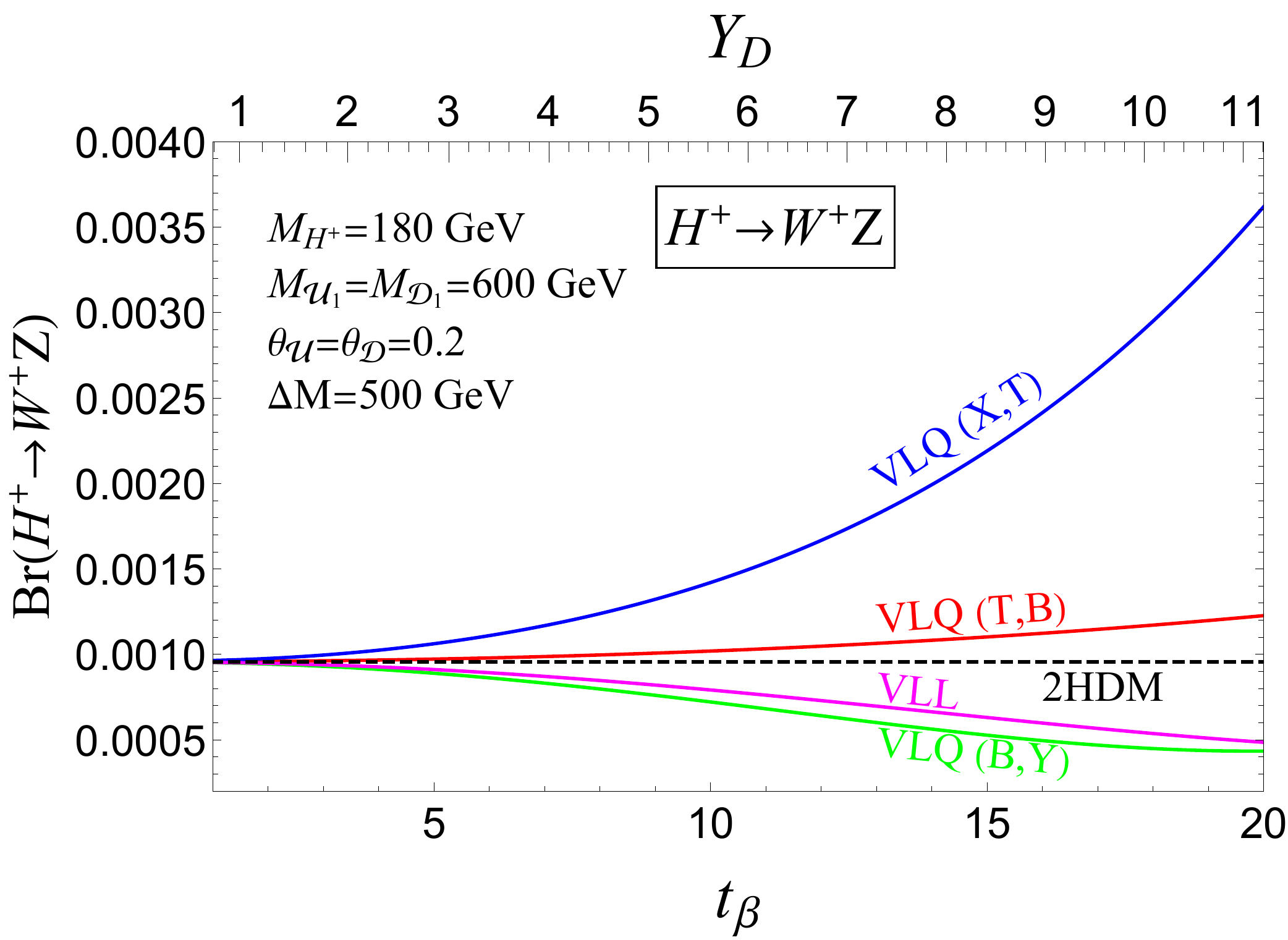}
\caption{
\baselineskip 3.5ex
$\br(H^+\to W^+ Z)$ as a function of $\Delta M$ for the fixed $\tb=5$ (left panels)
and $\tb$ for the fixed $\Dt M = 500\gev$ (right panels).
We set $\mch=180\gev$,
$M_{\mcu_1} = M_{\mcd_1}=600 \gev$ for the VLQs,
$M_{\mcu_1} = M_{\mcd_1}=300 \gev$ for the VLLs,
and $\theta_\mcu=\theta_\mcd=0.2$.
The numbers in the parenthesis denote the electric charges of VLQ.
The dashed lines represent the result in type-I 2HDM without the VLFs.
}
\label{fig:BRWZ}
\end{figure}

In Fig.~\ref{fig:BRWZ}, we show that the branching ratios of $H^\pm \to W^\pm Z$
as a function of $\Delta M$ for the fixed $\tb=5$ (left panels) and $\tb$ for the fixed $\Dt M = 500\gev$ (right panels).
We take $\mch=180\gev$, $M_{\mcu_1} = M_{\mcd_1}=600 \gev$ for the VLQs,
$M_{\mcu_1} = M_{\mcd_1}=300 \gev$ for the VLLs,
and $\theta_\mcu=\theta_\mcd=0.2$.
The VLQ loop contribution to $H^\pm \to W^\pm Z$
is not as large as that to $H^\pm \to W^\pm  \gm $,
typically a few tens of percent for $Y_D \simeq 5$.
We find that there is a strong correlation between $\br(H^\pm \to W^\pm Z)$
and the electroweak oblique parameter $\hat T$.
Our ansatz which guarantees $\hat T=0$ suppresses new contributions to $\br(H^\pm \to W^\pm Z)$.

\begin{figure}[h] \centering
\includegraphics[width=0.49\textwidth]{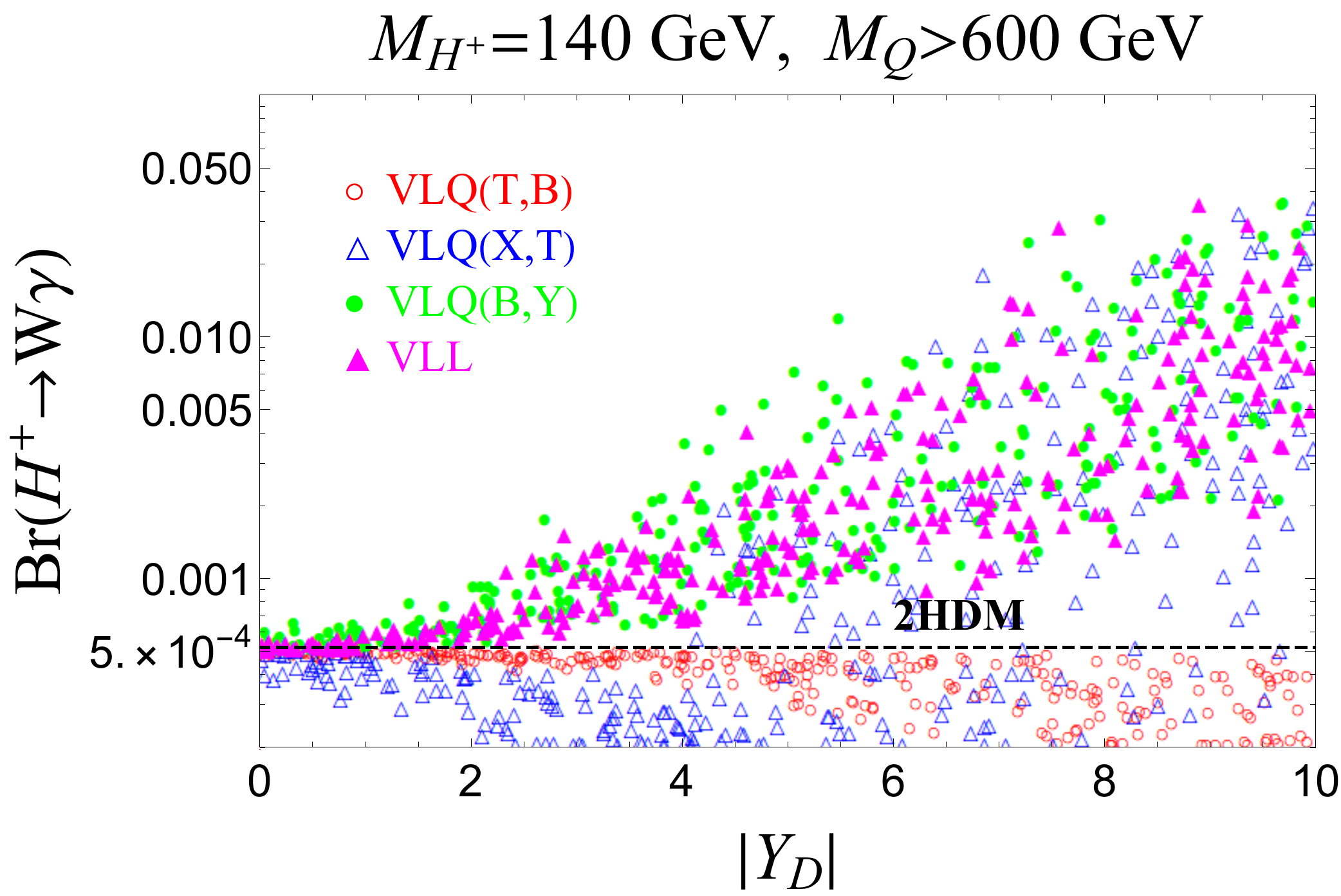}
\includegraphics[width=0.49\textwidth]{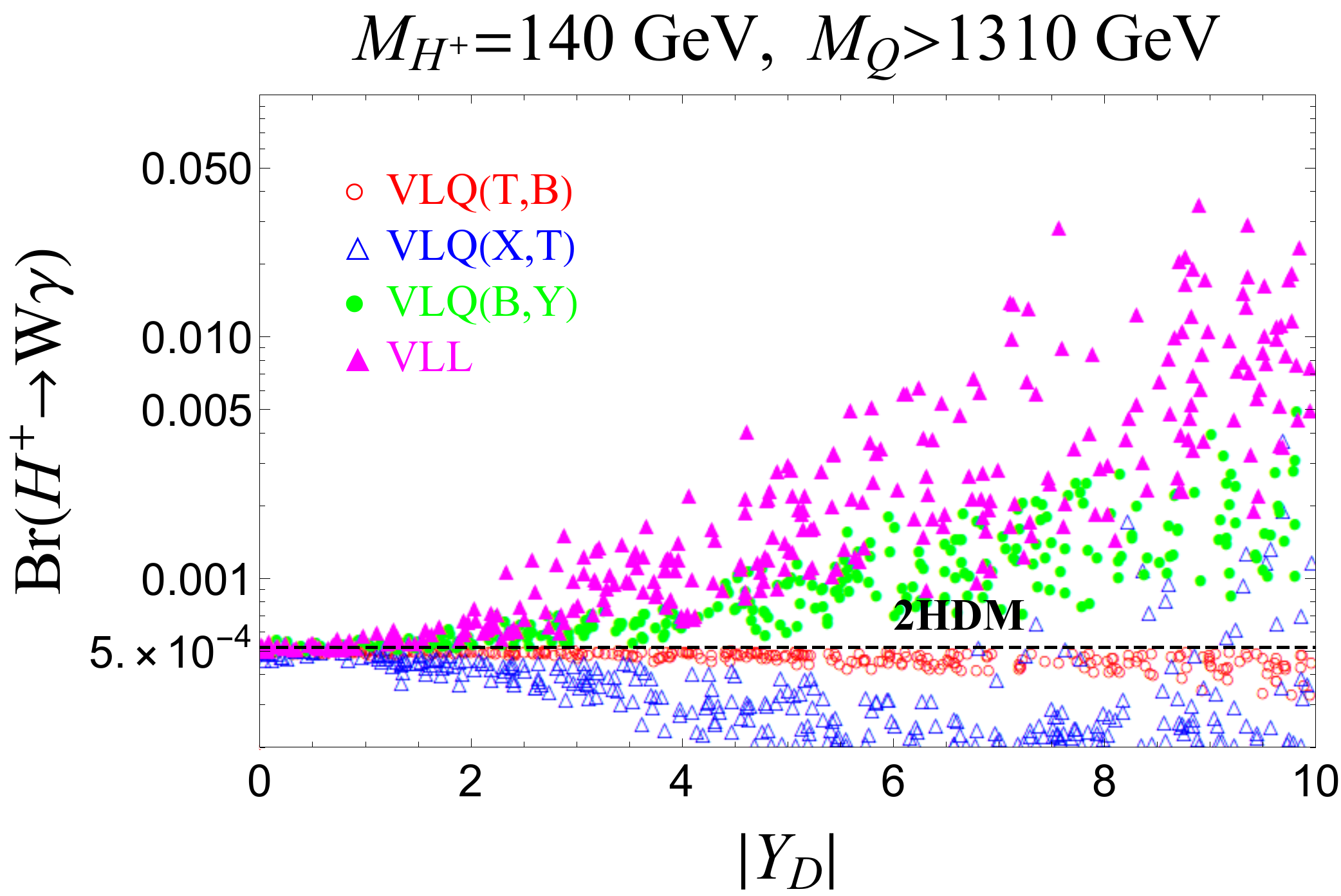}
\caption{
\baselineskip 3.5ex
Scatter plots for  $\br(H^+\to W\gamma)$ from the parameter sets which satisfy all of the
experimental constraints.
We set $\mch=140\gev$.
Dashed lines represent the results of the 2HDM without the VLFs.
}
\label{fig:scatter}
\end{figure}

The reader may question
whether the large enhancement of
$\br(H^\pm \to W^\pm\gm)$ happens only in the benchmark scenario.
To answer the question,
we 
take the $\mch=140\gev$ case and 
scan all of the parameters
in the range of
\begin{eqnarray}
\label{eq:scan:range}
&& M_{\mcf_1} < M_{\mcf_2}  \subset   [600, 3000]\gev ~~~(\textrm{for the VLQ with low mass})\,, \\ \nn
&& M_{\mcf_1} < M_{\mcf_2}  \subset   [1310, 5000]\gev ~~(\textrm{for the VLQ with high mass})\,, \\ \nn
&& M_{\mcf_1} < M_{\mcf_2}  \subset   [300, 1500]\gev ~~(\textrm{for the VLL})\,, \\ \nn
&& \tb \subset [1,50], \quad \theta_\mcu,\theta_\mcd \subset \left[ -{\pi}/{2}, {\pi}/{2}\right]\,.
\end{eqnarray}
Note that we independently span $\theta_\mcu$ and $\theta_\mcd$, not imposing
the condition of $\theta_\mcu = \theta_\mcd$.
The parameters in Eq.~(\ref{eq:scan:range}) determine
$Y_\mcu$ and $Y_\mcd$ through Eq.~(\ref{eq:mixingangle}).
Then we select the parameter sets that satisfy the constraints
from the Higgs precision data on $\kp_g$,
the upper bound on $\br(t \to b H^\pm)\times \br(H^\pm \to \tau^\pm\nu)$,
the electroweak oblique parameter $\hat{T}$,
and the perturbativity $|Y_{\mcu,\mcd}|<4 \pi$.
For the surviving parameter sets,
we show the scatter plots of $\br(H^+\to W\gamma)$ as a function of $|Y_\mcd|$
for the VLQs with low masses (left panel) and with high masses (right panel) in Fig.~\ref{fig:scatter}.
It is true that the benchmark scenario in Eq.~(\ref{eq:benchmark})
yields very large $\br(H^\pm\to W^\pm\gm)$, though not the maximum.
Nonetheless,
considerable parameter sets for the $(X,T)$,  $(B,Y)$,
and $(N,E)$ cases
allow at least one order of magnitude enhancement of $\br(H^\pm\to W^\pm\gm)$
for $|Y_\mcd| \gsim 5$.
It is fair to say that the VLFs in our model greatly enhance the branching ratio of $H^\pm\to W^\pm\gm$.
We caution the reader that the benchmark point for the $(X,T)$ case
does not represent the whole parameter space:
even for large $\left|Y_\mcd \right|\gsim 5$,
the $(X,T)$ contribution to
$\br(H^\pm\to W^\pm\gm)$ can be very destructive or very constructive,
while the benchmark point always enhances the branching ratio.
For heavier VLQ masses (right panel),
the range of the scatter plot is not as wide as that for low VLQ masses.
The scatter ranges  of the $(X,T)$, $(T,B)$, and $(B,Y)$ cases are quite separated.

\section{$H^\pm \to W^\pm \gamma$ mode at the LHC}
\label{sec:production}

At the LHC, the charged Higgs boson in the 2HDM
is produced in two ways, through the SM particles or through the resonant decay of $H$ or $A$.
The first category includes
\begin{eqnarray}
\label{eq:production:Nres}
 gg\to H^\pm W^\mp \bb, \quad  q \bar q \to H^+ H^- \,.
\end{eqnarray}
The process $gg\to H^+ W^- \bb$ for $\mch \ll m_t$ is the same as $gg \to \ttop$ followed by $t \to H^+ b$.
For $\mch\gg m_t$, it is effectively $g \bar{b} \to \bar{t} H^+$.
For the intermediate-mass $H^\pm$ ($140 \lsim \mch \lsim 200\gev$),
the full process $gg\to H^\pm W^\mp \bb$ 
at NLO should be considered because of the non-negligible 
effects from finite top-width, the significant interference between non-resonant and top-resonant diagrams,
and the sizable $K$-factor ($K \simeq 1.5$)~\cite{Degrande:2016hyf}.
Other production processes such as $q\bar q^\prime \to H^+ h/H^+ H$, $c \bar s \to H^+$, and $b\bar b \to W^- H^+$
have very small cross section,
one order of magnitude smaller than those in Eq.~(\ref{eq:production:Nres}).
Note that all of these production processes occur at tree level:
the VLFs do not play a role here.

\begin{figure}[h] \centering
\includegraphics[width=0.6\textwidth]{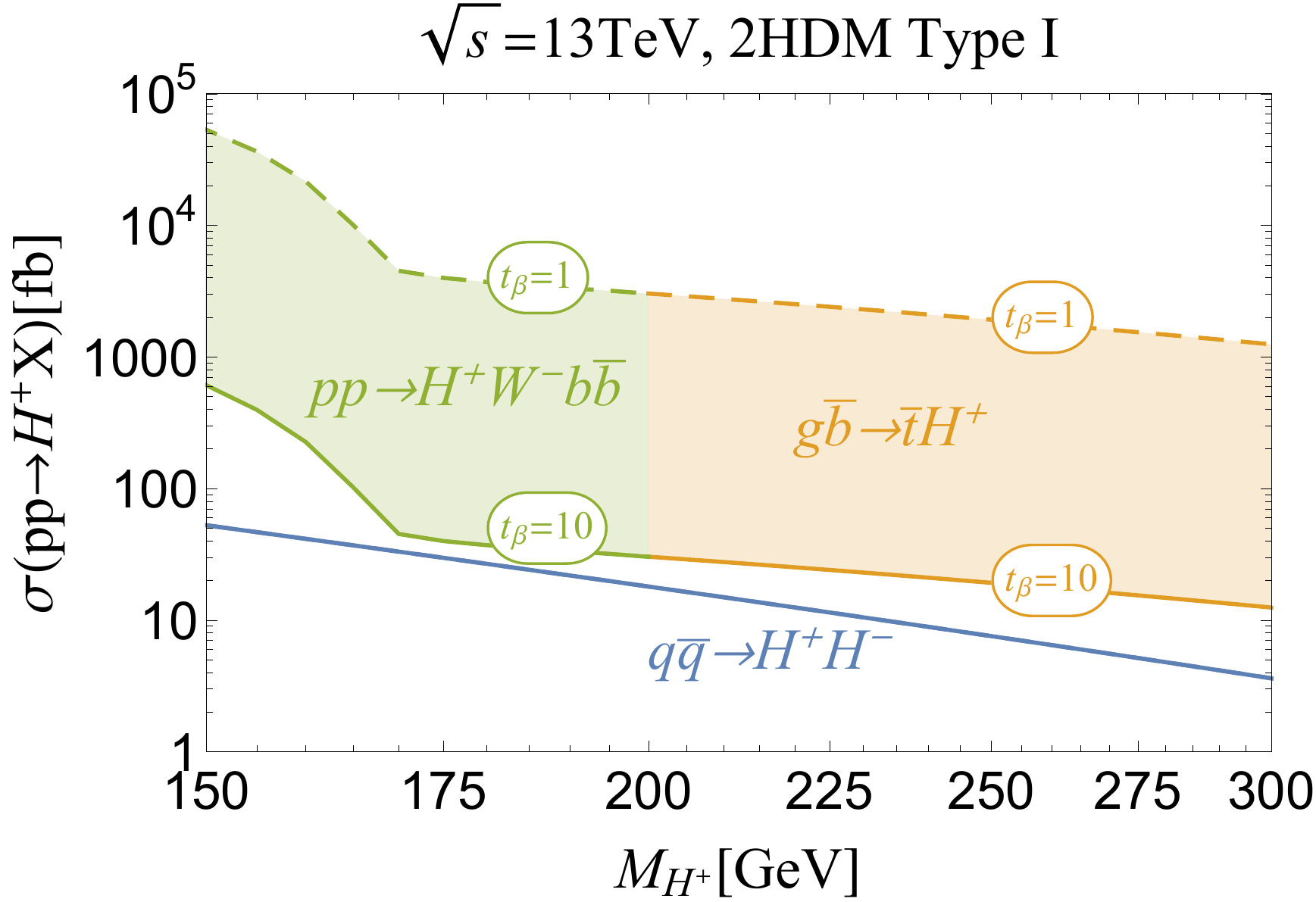}
\caption{
\baselineskip 3.5ex
Production cross sections of the charged Higgs boson as a function of its mass at the 13 TeV LHC.
We consider two cases of $\tb=1$ (dashed line) and $\tb=10$ (solid line).
}
\label{fig:production1}
\end{figure}

In Fig.~\ref{fig:production1}, we show the cross sections
of the production channels in Eq.~(\ref{eq:production:Nres}) as a function of $\mch$ at the 13 TeV LHC.
For $\mch \leq 200\gev$,
we use the full $gg\to H^\pm W^\mp \bb$ for type-I 2HDM~\cite{Degrande:2016hyf}.
For $\mch \geq 200\gev$,
the production process of $g\bar b \to \bar t H^+$~\cite{Kidonakis:2004ib,Zhu:2001nt,Plehn:2002vy,Berger:2003sm}
is presented, by using NNPDF~\cite{Ball:2014uwa} for the parton distribution function inside the proton.
We consider two $\tb$ cases,
$\tb=1$ (dashed line) and $\tb=10$ (solid line).
The production cross section, inversely proportional to $\tb^2$,
decreases with increasing $\tb$,
which is opposite to $\br(H^\pm \to W^\pm\gm)$.
The pair production $q\bar{q} \to H^+ H^-$~\cite{Willenbrock:1986ry}
via $s$-channel diagrams mediated by $\gm$ and $Z$
is independent of $\tb$.
The production cross section of $pp \to H^+ H^-$
is very small in the whole range of $\mch$,
being $\mco(1)\sim \mco(10)\fb$.

\begin{figure}[h!] \centering
\includegraphics[width=0.7\textwidth]{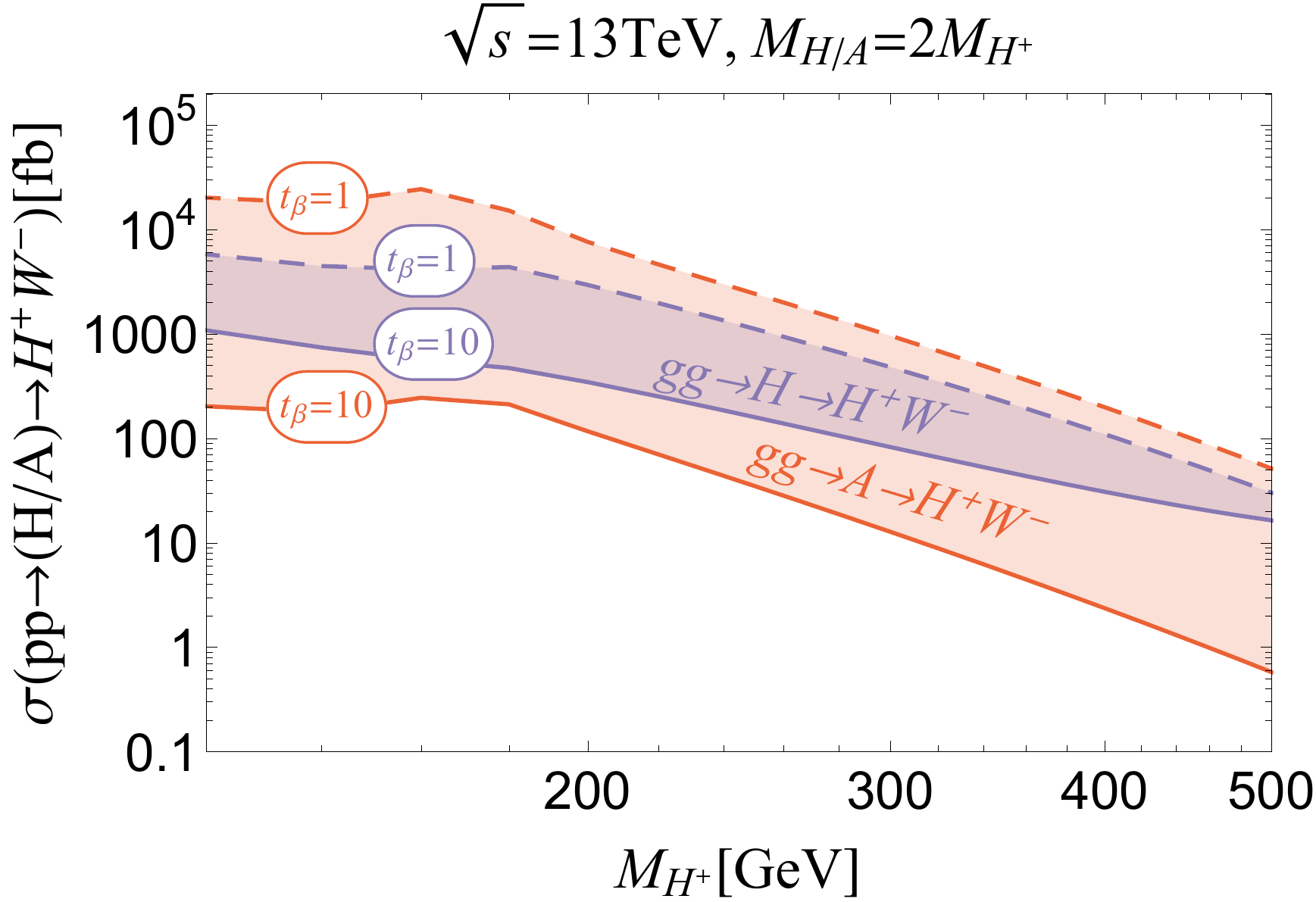}
\caption{\baselineskip 3.5ex
Cross sections of the resonance production channels for the charged Higgs boson
at the LHC as a function of its mass.
We use the {HIGLU} Fortran package \cite{Spira:1995mt} for estimating NNLO K-factors for neutral Higgs production.
We set $M_{\mcu_1} = M_{\mcd_1} = 600\gev$, $M_{\mcu_2} = M_{\mcd_2} = 1200\gev$,
$\theta_\mcu = \theta_\mcd = 0.2$, and $M_{H/A} = 2M_{H^+}$.
We consider two cases of $\tb=1$ (dashed line) and $\tb=10$ (solid line).
}
\label{fig:production2}
\end{figure}

Another way to produce the charged Higgs boson at the LHC
is through the resonant decay of other heavy Higgs bosons\footnote{
The process $gg\to H \to H^+ H^-$, which has more model-dependence, is not considered here.}:
\begin{eqnarray}
\label{eq:production2}
 gg\to H/A \to H^\pm W^\mp \,.
\end{eqnarray}
Note that the VLF contributions to the gluon fusion production of $H$ or $A$
are essentially negligible:
(i) the scattering amplitude of $gg \to A$
is proportional to the axial-vector coupling of the fermion in the loop,
which vanishes for the VLFs;
(ii) the VLF effects on the production process $gg \to H$
are very small because of 
the relation $y^H_{\mcf_1 \mcf_1}=-y^H_{\mcf_2 \mcf_2}$.
The decays of $H \to H^\pm W^\mp$ and $A \to H^\pm W^\mp$
occur from the following Lagrangian terms:
\begin{eqnarray}
{\cal L} \supset i  \frac{e}{2s_W} W_{\mu}^-\Big[ s_{\beta-\alpha}\Big(
H^+ \partial^\mu H - H \partial ^\mu H^+ \Big)
-i\Big( H^+ \partial^\mu A - A \partial ^\mu H^+ \Big) \Big]
\,.
\end{eqnarray}

Fig.~\ref{fig:production2} shows the production cross section of $gg \to H/A\to H^+ W^-$
as a function of $\mch$.
We set $M_{H/A} = 2M_{H^+}$,
$M_{\mcu_1} = M_{\mcd_1} = 600\gev$, $\Dt M = 600\gev$, and
$\theta_\mcu = \theta_\mcd = 0.2$ for two cases of $\tb=1$ (dashed line) and $\tb = 10$ (solid line).
Both $gg \to H \to H^+ W^-$
and $gg \to A \to H^+ W^-$
have sizable cross section of
$\mco(1)\sim\mco(10)\pb$ for the intermediate-mass charged Higgs boson.
For $\tb=1$,
$gg \to A \to H^+ W^-$ is more dominant
while for $\tb=10$, $gg \to H \to H^+ W^-$ is more important.
A crucial factor is the unknown parameter $M_{H/A}$,
which is set to be $2 \mch$ in Fig.~\ref{fig:production2}.
With increasing $M_{H/A}$,
$\sg(gg\to H/A)$
drops quickly:
if we double $M_{H/A}$,
$\sg(gg\to H)$ is about 30\% of that with $M_H=2 \mch$
and
$\sg(gg\to A)$ is only 10\%.
On the other hand,  $\br(H/A \to H^+ W^-)$
decreases only a few percent.

\begin{figure}[h!] \centering
\includegraphics[width=0.6\textwidth]{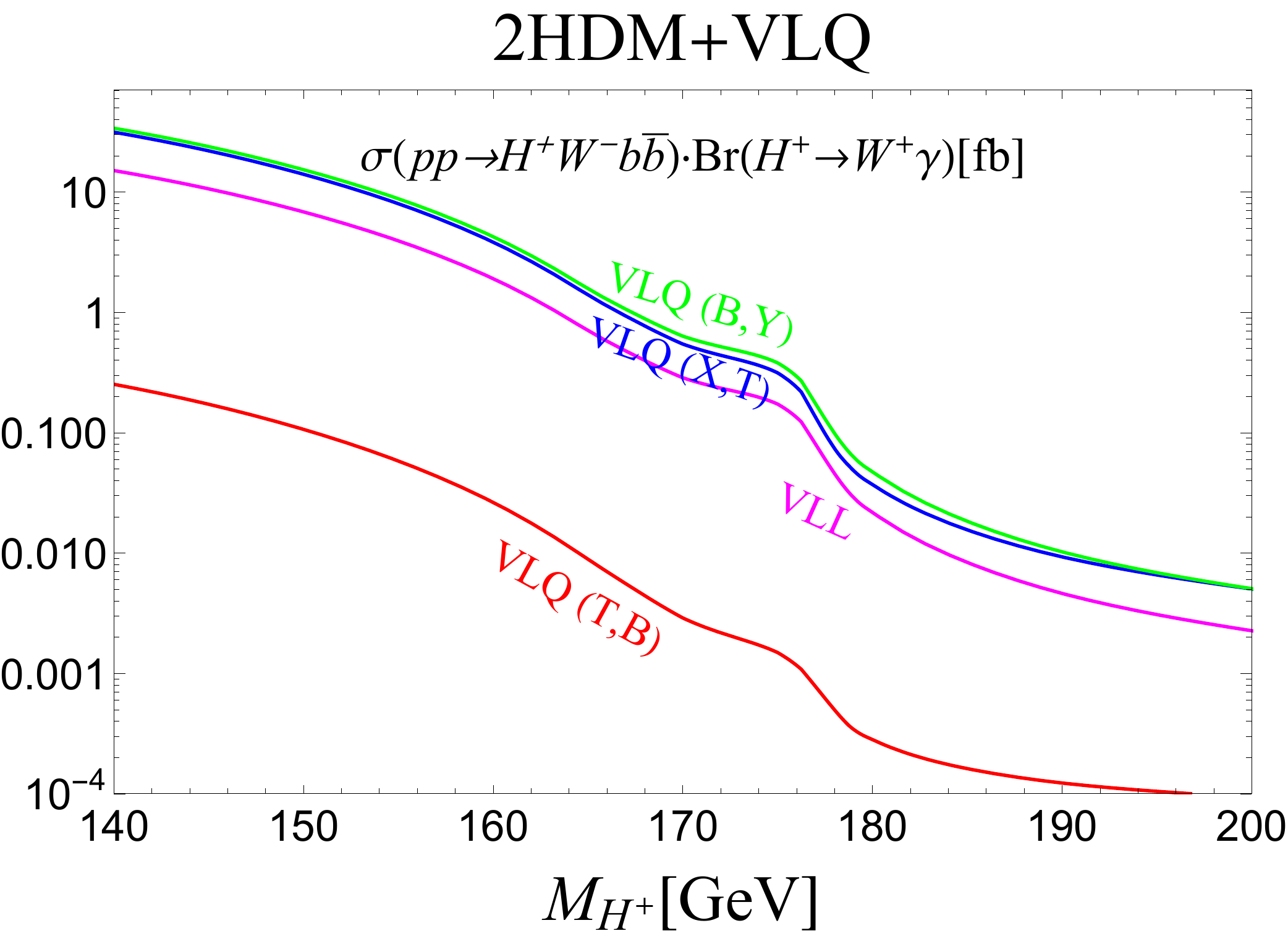}
\caption{\baselineskip 3.5ex
$\sg( pp \to H^+ W^- \bb )\times \br(H^+ \to W^+ \gm)$
as a function of $\mch$
at the LHC with $\sqs=13\tev$.
We set $\tb=10$, $M_{\mcu_1}=M_{\mcd_1}=600\gev$ for the VLQs, $M_{\mcu_1}=M_{\mcd_1}=300\gev$ for the VLLs, $\Dt M=600\gev$, and $\theta_\mcu=\theta_\mcd=0.2$.
}
\label{fig-SignalXSMHp1-tb10}
\end{figure}

In Fig.~\ref{fig-SignalXSMHp1-tb10}, we show
the signal rate $\sg( pp \to H^+ W^- \bb )\times \br(H^+ \to W^+ \gm)$
as a function of $\mch$
at the LHC with $\sqs=13\tev$. 
We set $\tb=10$, $M_{\mcu_1}=M_{\mcd_1}=600\gev$ for the VLQs, $M_{\mcu_1}=M_{\mcd_1}=300\gev$ for the VLLs,
$\Dt M=600\gev$, and $\theta_\mcu=\theta_\mcd=0.2$.
The common feature is that $\sg\times\br (W\gm)$ drops fast with increasing $\mch$,
especially after the $tb$ threshold.
In the mass range of $\mch \lsim 160\gev$,
the $W\gm$ mode 
has a sizable $\sg\times \br$  at the LHC,
except for the $(T,B)$ case.

\begin{figure}[h!] \centering
\includegraphics[width=0.47\textwidth]{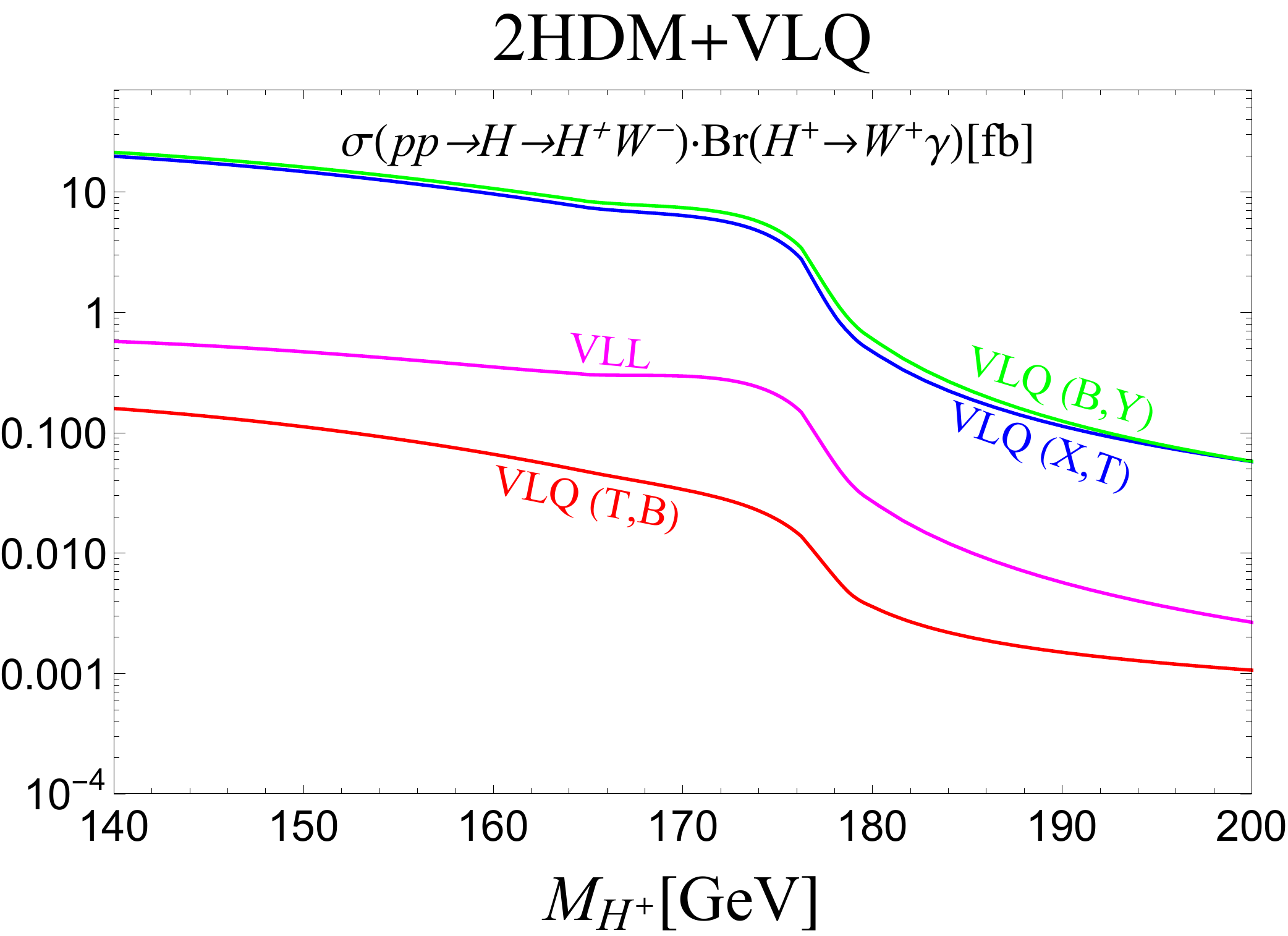}
\includegraphics[width=0.47\textwidth]{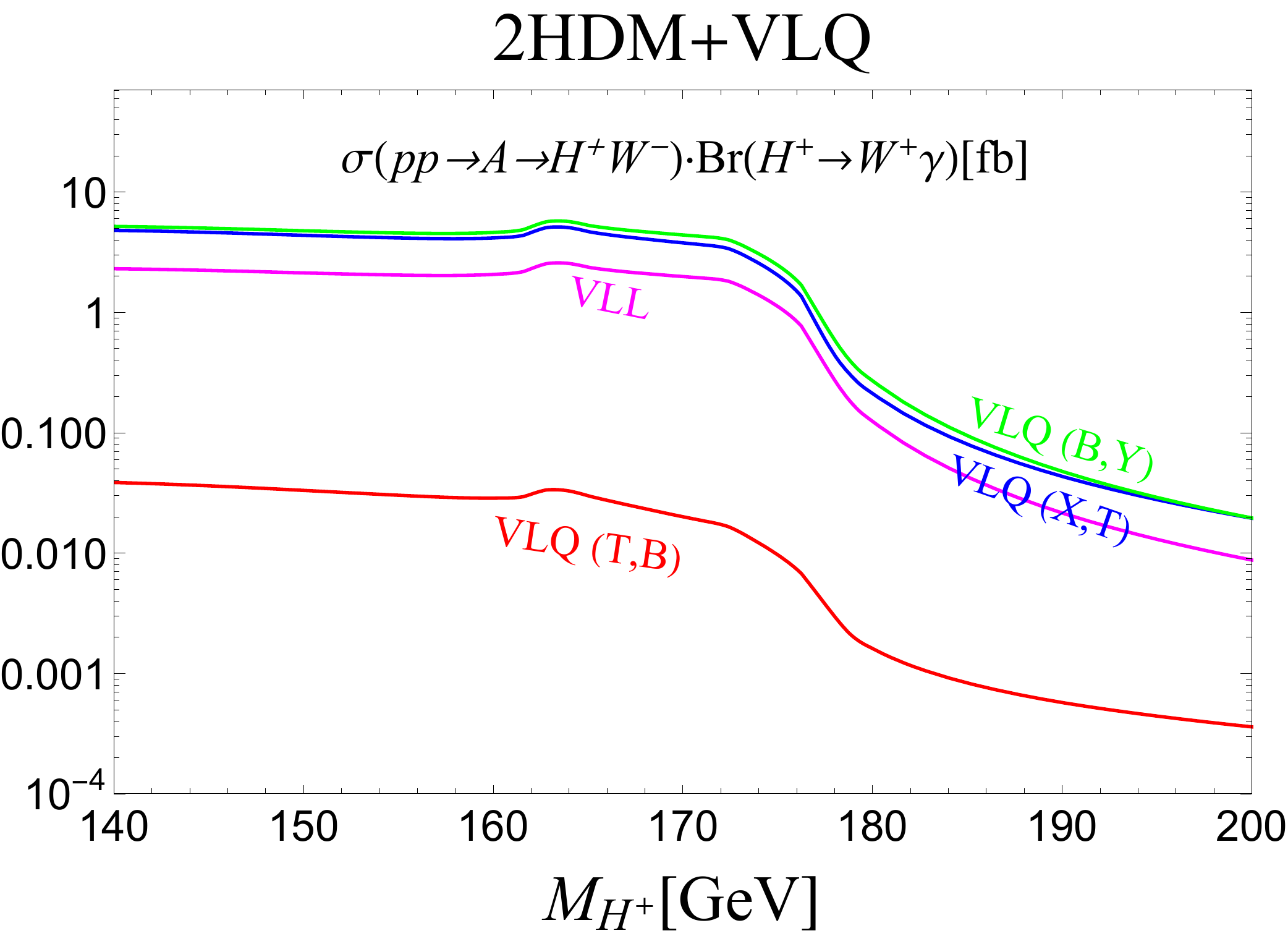}
\caption{\baselineskip 3.5ex
$\sg( gg \to H\to H^+ W^-)\times \br(H^+ \to W^+ \gm)$ (left panel)
and $\sg( gg \to A \to H^+ W^-)\times \br(H^+ \to W^+ \gm)$ (right panel)
as a function of $\mch$
at the 13 TeV LHC.
We set $\tb=10$, $M_H=M_A=2 \mch$, $M_{\mcu_1}=M_{\mcd_1}=600\gev$ for the VLQs, $M_{\mcu_1}=M_{\mcd_1}=300\gev$ for the VLLs,  $\Dt M=600\gev$, and $\theta_\mcu=\theta_\mcd=0.2$.
}
\label{fig-SignalXSMHp2-tb10}
\end{figure}

Fig.~\ref{fig-SignalXSMHp2-tb10} presents $\sg( gg \to H\to H^+ W^-)\times \br(H^+ \to W^+ \gm)$ (left panel)
and $\sg( gg \to A \to H^+ W^-)\times \br(H^+ \to W^+ \gm)$ (right panel)
as a function of $\mch$
at the 13 TeV LHC.
We set $\tb=10$, $M_H=M_A=2 \mch$, 
$M_{\mcu_1}=M_{\mcd_1}=600\gev$ for the VLQs, $M_{\mcu_1}=M_{\mcd_1}=300\gev$ for the VLLs, $\Dt M=600\gev$,
and $\theta_\mcu=\theta_\mcd=0.2$.
As a resonance production process,
the drop in the signal rate $\sg \times\br$ with increasing $\mch$
is not as much as in the $pp \to H^+ W^- \bb $ process.
The $(X,T)$ and $(B,Y)$ cases have $\sg\times \br \gsim 1\fb$ before the $tb$ threshold.

\begin{figure}[h!] \centering
\includegraphics[width=0.56\textwidth]{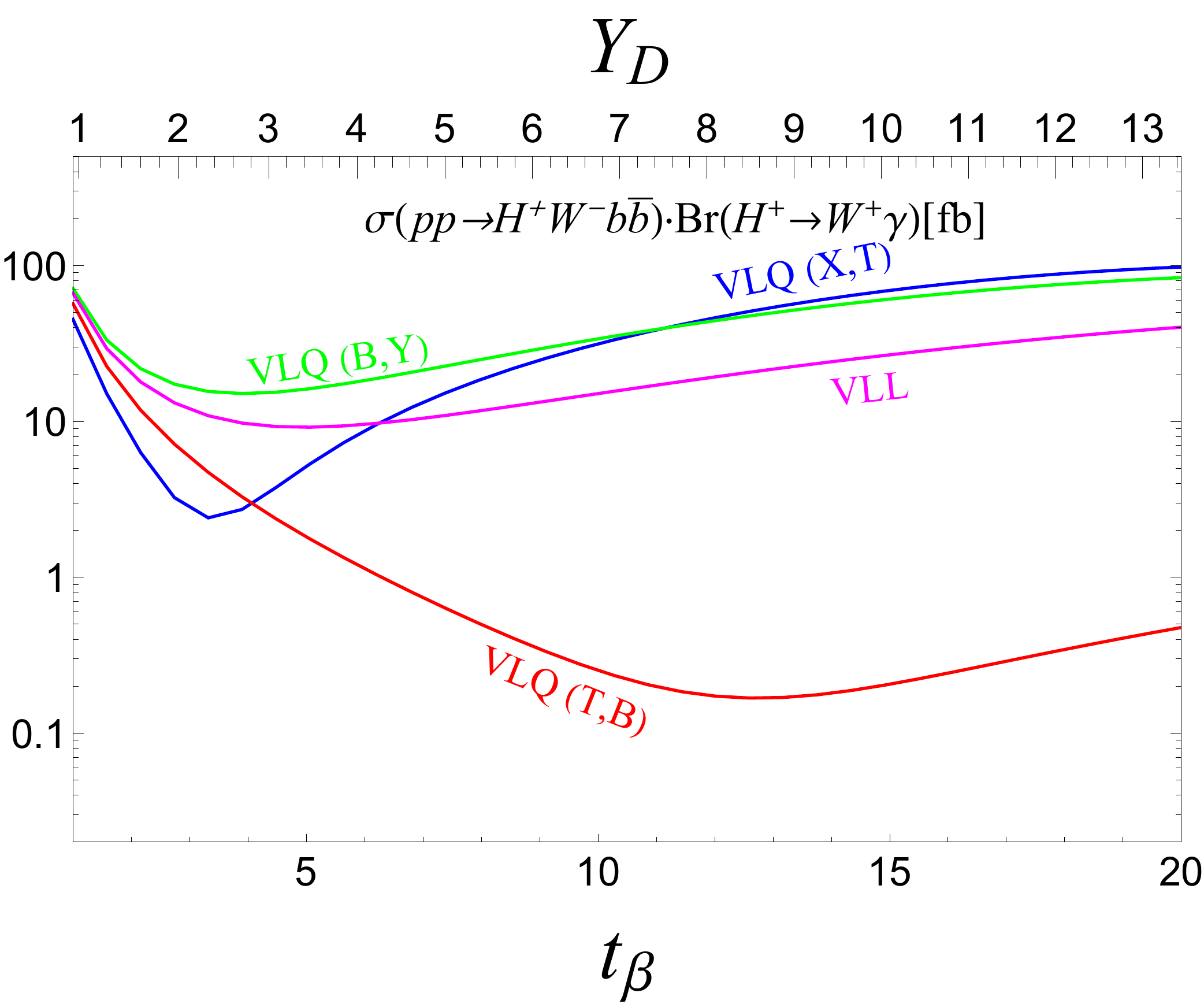}
\caption{\baselineskip 3.5ex
$\sg( pp \to H^+ W^- \bb )\times \br(H^+ \to W^+ \gm)$
as a function of $\tb$
at the LHC with $\sqs=13\tev$.
We set $\mch=140\gev$, $M_{\mcu_1}=M_{\mcd_1}=600\gev$ for the VLQs, 
$M_{\mcu_1}=M_{\mcd_1}=300\gev$ for the VLL, $\Dt M=600\gev$, and $\theta_\mcu=\theta_\mcd=0.2$.
}
\label{fig-SignalXStb-single-top}
\end{figure}

Now we further investigate the dependence of the model parameters on the signal rate.
The most crucial one is $\tb$, because of decreasing $\sg(pp \to H^\pm X)$ 
but increasing $\br(H^\pm \to W^\pm \gm)$
with increasing $\tb$.
In Fig.~\ref{fig-SignalXStb-single-top},
we show $\sg( pp \to H^+ W^- \bb )\times \br(H^+ \to W^+ \gm)$
as a function of $\tb$
at the 13 TeV LHC.
We set $\mch=140\gev$, $M_{\mcu_1}=M_{\mcd_1}=600\gev$ for the VLQs,
$M_{\mcu_1}=M_{\mcd_1}=300\gev$ for the VLLs,  $\Dt M=600\gev$, and $\theta_\mcu=\theta_\mcd=0.2$.
All of the four VLF cases show similar behaviors of  $\sg\times\br$ about $\tb$.
Up to some critical value of $\tb$, 
$\sg\times\br$ decreases because of decreasing $\sg$ with increasing $\tb$. 
After some critical value of $\tb$,
$\sg\times\br$ increases with $\tb$
as the branching ratio increase is dominant.
It is remarkable that there exists a sizable portion of parameter $\tb$
which allows significant signal rate
for all of the VLF cases.
For the $(X,T)$, $(B,Y)$, and VLL cases,
$\sg\times\br \gsim 10\fb$ in the whole range of $\tb$.
The tricky $(T,B)$ case has a chance for the discovery in the small $\tb$ region:
$\sg\times\br \gsim 10 ~(1)\fb$ when $\tb \lsim 1.8 ~(6.2)$.

\begin{figure}[h!] \centering
\includegraphics[width=0.47\textwidth]{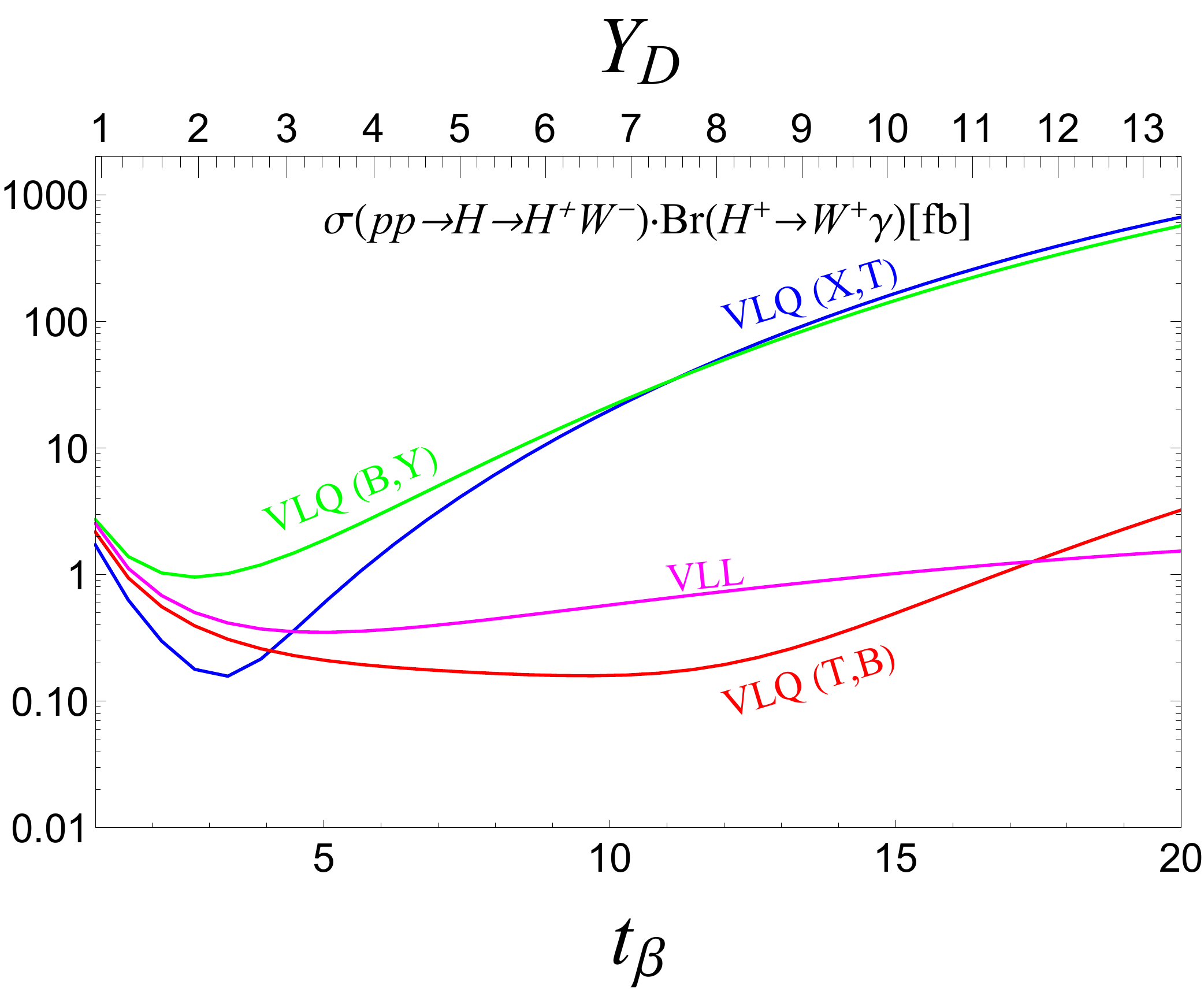}
\includegraphics[width=0.47\textwidth]{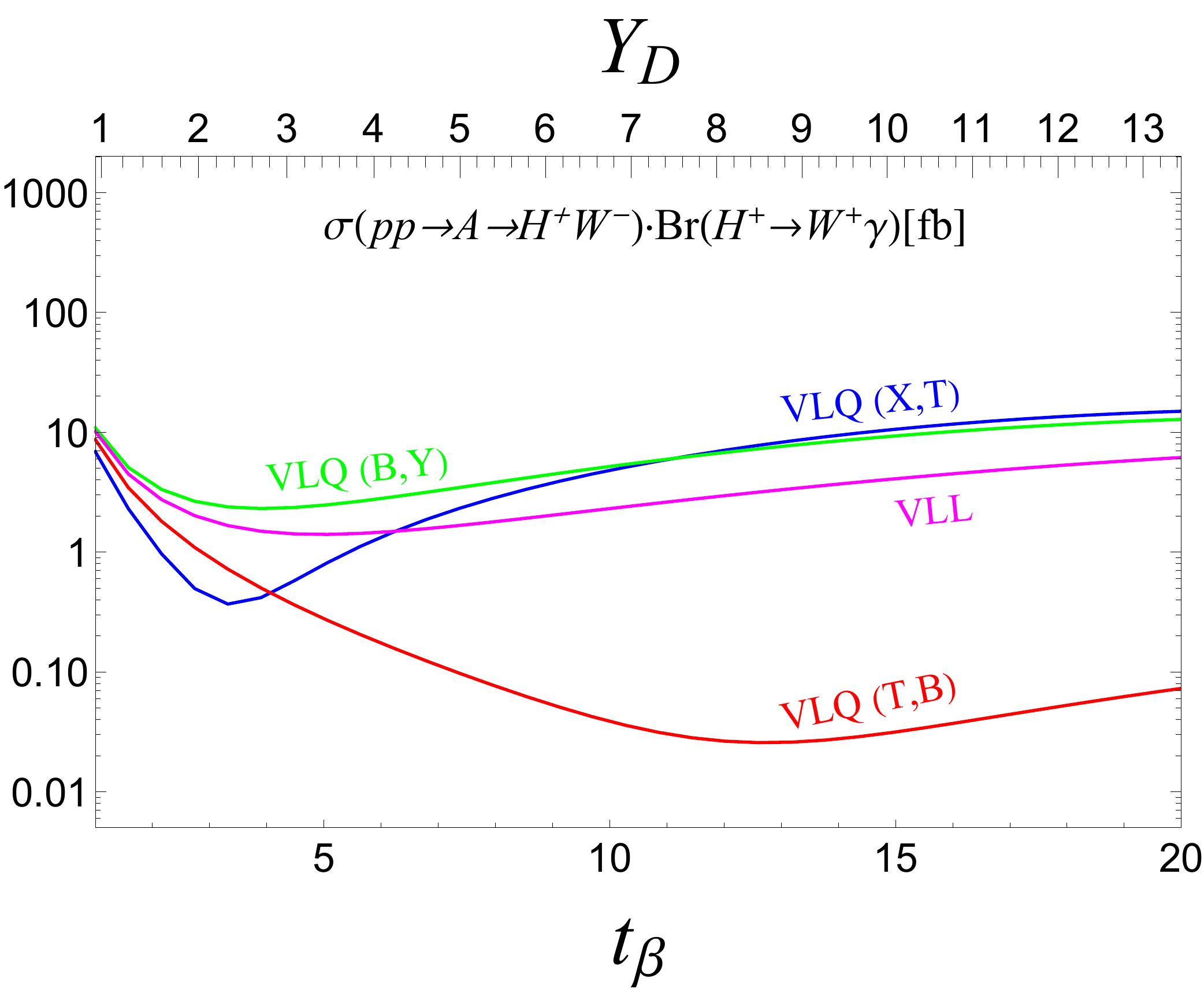}
\caption{\baselineskip 3.5ex
$\sg( gg \to H \to H^+ W^-)\times \br(H^+ \to W^+ \gm)$ (left panel)
and $\sg( gg \to A \to H^+ W^-)\times \br(H^+ \to W^+ \gm)$ (right panel)
as a function of $\tb$
at the 13 TeV LHC.
We set $\mch=140\gev$, $M_H=M_A=2 \mch$, $M_{\mcu_1}=M_{\mcd_1}=600\gev$ for the VLQs, $M_{\mcu_1}=M_{\mcd_1}=300\gev$ for the VLLs,  $\Dt M=600\gev$, and $\theta_\mcu=\theta_\mcd=0.2$.
}
\label{fig-SignalXStb-H-A}
\end{figure}

Next we present $\sg(gg \to H/A \to H^+ W^-)\times \br(H^+ \to W^+ \gm)$
as a function of $\tb$ in Fig.~\ref{fig-SignalXStb-H-A},
the $H$-resonance one (left panel) and the $A$-resonance one (right panel).
We set $\mch=140\gev$, $M_H=M_A=2 \mch$, $M_{\mcu_1}=M_{\mcd_1}=600\gev$ for the VLQs,
$M_{\mcu_1}=M_{\mcd_1}=300\gev$ for the VLLs, $\Dt M=600\gev$, and $\theta_\mcu=\theta_\mcd=0.2$.
Since the suppression of the production cross section
by large $\tb$ is weak for the $H$ resonance as shown in Fig.~\ref{fig:production2},
the increase of $\sg\times\br$ with respect to $\tb$ is much larger for the $gg\to H$ production channel.
Through the $H$-radiation, $\sg\times\br$ can excess $\sim 10\fb$ if $\tb \gsim 8$:
even $100\fb$ is possible if $\tb \gsim 13$. 
The $A$ resonant production of the charged Higgs boson is also very sizable  for the $(X,T)$ and $(B,Y)$ cases.
$\sg\times\br \simeq 10\fb$ can be achieved for $\tb \gsim 12$.

\section{Conclusions}
\label{sec:conclusions}
Targeting the intermediated-mass charged Higgs boson $H^\pm$,
we have explored the theoretical possibility that 
the branching ratios of its radiative decays into $W^\pm \gm$ and $W^\pm Z$
are large enough for the LHC discovery. 
We considered a two-Higgs-doublet model
with a vectorlike fermion (VLF) $SU(2)$ doublet $\mcq$ and two singlets $\mcu$ and $\mcd$.
Various VLF cases with different electric charges have been studied, including
the vectorlike lepton $(N,E)$ with the electric charge $(0,-1)$
as well as the vectorlike quarks $(X,T)$ with $(5/3,2/3)$,
$(T,B)$ with $(2/3,-1/3)$,
and $(B,Y)$ with $(-1/3,-4/3)$.
For the large enhancement of the loop-induced decays,
we suggest the type-I-II 2HDM
where the SM fermions are assigned in type-I while the VLFs are in type-II.

Introducing a VLF doublet  and two singlets,
which is necessary for the interaction with the Higgs doublet fields,
plays a crucial role.
As being vectorlike,
one generation of the new fermions has two up-type fermions, $\mcu_1$ and $\mcu_2$,
and two down-type fermions, $\mcd_1$ and $\mcd_2$.
This new fermion spectrum allows significant cancellation among different VLF contributions
to the Higgs precision data as well as to the electroweak oblique parameters, especially $\hat T$.
Sizable cancellation to the $h$-$g$-$g$ vertex
occurs naturally because the $h \mcf_1 \mcf_1 $ coupling is opposite to $h \mcf_2 \mcf_2 $.
The cancellation for the electroweak oblique parameter $\hat T$
requires some fine-tuning.
We proposed an ansatz to ensure $\hat T=0$ such that
$M_{\mcu_1}=M_{\mcd_1}$, $M_{\mcu_2}=M_{\mcd_2}$, and $\theta_\mcu=\theta_\mcd$,
which is not so artificial.
We have also included the constraints from direct search bounds on the VLFs and charged Higgs boson
at the LHC.

We presented the loop-induced amplitudes of $H^\pm \to W^\pm \gm$ and $H^\pm \to W^\pm Z$
from the VLFs as well as the SM $t$ and $b$ quarks.
The branching ratios of two radiative decays
as a function of $\mch$ show that the $W\gm$  mode is
very efficient for the mass range of $[110,170]\gev$
and the $WZ$ mode is good for $\mch\simeq m_t$. 
We found that $\br(H^\pm \to W^\pm Z)$
is not changed much by the VLF contributions,
because of the strong correlation with the electroweak oblique parameter $\hat T$.
On the other hand, the $W\gm$ mode can be greatly enhanced
for large $\Dt M (=M_{\mcu_2}-M_{\mcu_1})$ and large $\tb$:
even two orders of magnitude increase is possible.
In the details, the 
four VLF cases show different behaviors.
In the $(B,Y)$ and $(N,E)$ cases, the branching ratio
is enhanced in the whole parameter region.
In the $(X,T)$ and $(T,B)$ cases, however,
the branching ratio for moderate $\Dt M$ and $\tb$ is smaller than that without the VLFs
because of 
the destructive interference with the top and bottom quark contributions.
For large $\Dt M$ and $\tb$,
the new physics contributions win the SM ones,
enhancing the branching ratio.
But the $(T,B)$ case requires very large $\Dt M$ or $\tb$ for positive contribution,
which endangers the perturbativity of the down-type Yukawa coupling.

We have also studied the production of the charged Higgs boson at the LHC,
through the SM particles, $pp\to H^\pm W^\mp \bb$,
and the resonant decay of a heavy Higgs boson $H$ or $A$, $gg \to H/A \to H^\pm W^\mp$.
The production cross sections decrease with increasing $\mch$,
more rapidly for $pp\to H^\pm W^\mp \bb$.
The signal rate $\sg \times \br(H^\pm \to W^\pm\gm)$ at the 13 TeV LHC
was also calculated.
In a large portion of the parameter space,
$\sg\times\br$ for the intermediate-mass charged Higgs boson exceeds 10 fb.

In conclusion,
the radiative decay mode $W\gm$
can serve as an alternative channel to probe
the intermediate-mass charged Higgs boson.
A theoretically viable model
in the extended type-I 2HDM
with the vectorlike fermions
was suggested to allow the
great enhancement of the $W\gm$ branching ratio.
We expect that this study helps the LHC to search for the charged Higgs boson.

\acknowledgments
We thank Eung Jin Chun for the helpful comments.
The work of J.S.
was supported by the National Research Foundation of Korea, Grant No. NRF-2016R1D1A1B03932102.
Y.W.Y was supported by Basic Science Research Program through the National Research Foundation of Korea,
Grant No. 2017R1A6A3A11036365.

\appendix

\section{Vacuum-polarization amplitudes of the SM gauge bosons}
\label{appendix:Pi}
For the electroweak oblique parameters $\hat{S}$, $Y$, and $W$,
we need the first and second derivatives of the transverse vacuum polarization amplitudes of the SM gauge bosons,
which are explicitly shown in Ref.~\cite{Cynolter:2008ea}.
However, we found some typos
in their results.
The correct ones are
\begin{eqnarray}
\label{eq:der:Pol}
\tilde \Pi_{V+A}^{\prime}(0) &=& \frac{1}{3}
\left[ {\rm Div} + \ln\Big(\frac{\mu^2}{m_a m_b}\Big)\right]
 + \frac{m_a^4-8m_a^2m_b^2+m_b^4}{9(m_a^2-m_b^2)^2}  \\ \nn
 && + \frac{(m_a^2+m_b^2)(m_a^4-4m_a^2 m_b^2+m_b^4)}{6(m_a^2-m_b^2)^3} \ln \Big(\frac{m_b^2}{m_a^2}\Big) \,, \\[5pt]
 \tilde \Pi_{V-A}^{\prime}(0) &=& m_a m_b \left[  \frac{(m_a^2 + m_b^2)}{2(m_a^2-m_b^2)^2}
 + \frac{m_a^2 m_b^2}{(m_a^2-m_b^2)^3} \ln \Big(\frac{m_b^2}{m_a^2}\Big)\right]\,,\\[5pt]
 \tilde \Pi_{V+A}^{\prime\prime}(0) &=&
 \frac{(m_a^2+m_b^2)(m_a^4-8m_a^2 m_b^2 +m_b^4)}{4(m_a^2-m_b^2)^4}
 - \frac{3m_a^4 m_b^4}{(m_a^2 - m_b^2)^5} \ln \Big(\frac{m_b^2}{m_a^2}\Big)\,, \\[5pt]
 \tilde \Pi_{V-A}^{\prime\prime}(0) &=&
 m_a m_b \left[ \frac{(m_a^4+ 10 m_a^2 m_b^2 + m_b^4 )}{3(m_a^2 - m_b^2)^4}
 + \frac{2(m_a^2+m_b^2) m_a^2 m_b^2}{(m_a^2-m_b^2)^5 } \ln \Big( \frac{m_b^2}{m_a^2} \Big)\right].
 \end{eqnarray}

In our type-I-II 2HDM, $\hat{S}$ is
\begin{eqnarray}
 \hat S &=& \frac{g^2 N_C}{16\pi^2 m_W^2} \bigg[
             ~ c_\mcu^2(2 Q_\mcu - c_\mcu^2) \tilde \Pi_V^{\prime}(M_{\mcu_1},M_{\mcu_1},0)
             + s_\mcu^2(2Q_\mcu - s_\mcu^2) \tilde \Pi_V^{\prime}(M_{\mcu_2},M_{\mcu_2},0) \nn\\
 &&\qquad\qquad - c_\mcd^2(2Q_\mcd + c_\mcd^2) \tilde \Pi_V^{\prime}(M_{\mcd_1},M_{\mcd_1},0)
             - s_\mcd^2(2Q_\mcd + s_\mcd^2) \tilde \Pi_V^{\prime}(M_{\mcd_2},M_{\mcd_2},0) \nn \\
 &&\qquad\qquad  -2s_\mcu^2 c_\mcu^2 \tilde \Pi_V^{\prime}(M_{\mcu_1},M_{\mcu_2},0)
             -2s_\mcd^2 c_\mcd^2 \tilde \Pi_V^{\prime}(M_{\mcd_1},M_{\mcd_2},0)
 \bigg] \,,
\end{eqnarray}
where $N_C =3~(1)$ for VLQ (VLL) is the color factor.
$Y$ and $W$ are
\begin{eqnarray}
  W &=&  \frac{g^2 m_W^2 N_C}{32\pi^2} \bigg[
               c_\mcu^4 \tilde \Pi_V^{\prime\prime}(M_{\mcu_1},M_{\mcu_1},0)
             + s_\mcu^4 \tilde \Pi_V^{\prime\prime}(M_{\mcu_2},M_{\mcu_2},0) \nn\\
 && \qquad\qquad + c_D^4 \tilde \Pi_V^{\prime\prime}(M_{\mcd_1},M_{\mcd_1},0)
             + s_\mcd^4 \tilde \Pi_V^{\prime\prime}(M_{\mcd_2},M_{\mcd_2},0) \nn\\
 && \qquad\qquad +2s_\mcu^2 c_\mcu^2 \tilde \Pi_V^{\prime\prime}(M_{\mcu_1},M_{\mcu_2},0)
             +2s_\mcd^2 c_D^2 \tilde \Pi_V^{\prime\prime}(M_{\mcd_1},M_{\mcd_2},0)
 \bigg] \,, \\
   Y &=& \Big(\frac{g^{\prime }}{g}\Big)^2 W \,.
 \end{eqnarray}

\section{Decay Form-Factors for $H^\pm \to W^\pm \gm/W^\pm Z$}
\label{appendix:form:factors}

\subsection{Loop function}

 For the one loop calculation, we express the result in term of the loop functions
of the LoopTools~\cite{Hahn:1998yk}.
Two point function defines $B_l$'s as
\begin{eqnarray}
 B_{0} \big(p^2, m_1^2, m_2^2\big) &\equiv& \frac{\mu^{4-D}}{i\pi^{D/2} r_\Gamma}
\int  d^D q \frac{1}
{\big[q^2-m_1^2\big]\big[(q+p)^2-m_2^2\big]}\,, \\
 p_\mu B_{1} \big(p^2, m_1^2, m_2^2\big) &\equiv& \frac{\mu^{4-D}}{i\pi^{D/2} r_\Gamma}
\int d^D q \frac{ q_\mu}
{\big[q^2-m_1^2\big]\big[(q+p)^2-m_2^2\big]}\,,
\end{eqnarray}
where $\mu$ is the renormalization scale, $D=4-2\es$, and $
r_\Gamma = {\Gamma^2(1-\epsilon)\Gamma(1+\epsilon)}/{\Gamma(1-2\epsilon)}$.

The tensorial integral for the one-loop three point function is defined by
\begin{eqnarray}
&& T^3_{\mu_1 \ldots \mu_P} \big(p_1^2, p_2^2, (p_1+p_2)^2, m_1^2, m_2^2, m_3^2\big) \nn\\
&&\qquad \equiv \frac{\mu^{4-D}}{i\pi^{D/2} r_\Gamma}
\int d^Dq \frac{q_{\mu_1}\ldots q_{\mu_P}}
{\big[q^2-m_1^2\big]\big[(q+p_1)^2-m_2^2\big]\big[(q+p_1+p_2)^2-m_3^2\big]}\,.
\end{eqnarray}
The decompositions of the tensorial integrals up to rank 2 are
\begin{eqnarray}
T^3 &=& C_0 \,,\nn\\
T^3_{\mu} &=& k_{1\mu} C_1 + k_{2\mu} C_2 \,, \nn \\
T^3_{\mu \nu} &=& g_{\mu\nu} C_{00} + \sum_{i,j=1}^2 k_{i\mu} k_{j\,\nu} C_{ij} \,,
\end{eqnarray}
where $k_1=p_1$ and $k_2 = p_1+p_2$.
All of
the coefficient functions of $B_i$, $C_i$ and $C_{ij}$ are numerically computed by {LoopTools}.
Note that $C_{00}$ and $B_i$ have UV divergence
which should be canceled out.

\subsection{Decay Form-Factors from the SM quark contributions}
We describe the form factors defined in Eq.~(\ref{eq:FF}) for each single diagram shown in Fig.~\ref{fig:diagrams}.
We compute the diagrams in the unitary gauge,
and use the dimensional regularization with $D=4-2\epsilon$ in the $\overline{\rm{MS}}$ scheme.
As for the UV divergence, we show only the $1/\epsilon$ term.
Since there is no tree-level coupling for the $H^+ W^- V$ vertex,
all of the UV divergences should be canceled out among themselves
 after summing all the diagrams.
This cancellation serves as a validation of the calculation.
For notational simplicity, we introduce the normalized gauge couplings and Yukawa couplings as
\bea
\label{eq:simple:couplings}
&&\gh^\gm_{ff} = s_W c_W Q_f, \quad \gh^Z_{WW} = c_W^2, \quad \gh^\gm_{WW}=-s_W c_W,
\\[5pt] \nn
&&\gh^{L}_{Zff} = T^3_f - Q_f s_W^2, \quad
\gh^{R}_{Zff} = -Q_f s_W^2
\\ \nn
&& y^L_{H^+tb} = \frac{ \sqrt{2} V_{tb} m_t}{v t_\beta},\quad
y^R_{H^+tb} = -\frac{ \sqrt{2} V_{tb} m_b}{v t_\beta}.
\eea

We first present the results for $H^+ \to W^+ \gamma$.
In the SM model,
the main contribution is from the top and bottom quarks.
 Since the decay involves a photon,
${\cal M}_1$ is determined by ${\cal M}_2$ through the Ward identity in Eq.~(\ref{eq:WardIdentity}).
We separately present the expressions of ${\cal M}_2$ and ${\cal M}_3$ from the diagrams (a), (b),
and (c) in Fig.~\ref{fig:diagrams}.
${\cal M}_2$'s are
\begin{eqnarray}
{\cal M}_{2}^{\rm (a)} &=& 0,
\\ \nn
\label{eq:FormFactor:SM:M2}
{\cal M}_{2}^{\rm (b)} &=&
2 \, \gh^\gm_{tt} \, \mch
\bigg[  y^L_{H^+tb}m_t(C_1-C_2-2C_{12}-2C_{22})
\\ \nn &&\qquad \qquad \qquad \qquad
- y^R_{H^+tb}m_b
 (C_0+C_1+3C_2+2C_{12}+2C_{22}) \bigg]
\,,  \\ \nn 
{\cal M}_{2}^{\rm (c)} &=&  {\cal M}_{2}^{\rm (b)}
\Big( y^L_{H^+tb} \leftrightarrow  y^R_{H^+tb}\,, t \leftrightarrow b \Big)
\,,
\end{eqnarray}
and ${\cal M}_{3}$'s are
\begin{eqnarray}
\label{eq:FormFactor:SM:M3}
{\cal M}_{3}^{\rm (a)} &=& 0,
\\ \nn
{\cal M}_{3}^{\rm (b)} &=& 2 \, \gh^\gm_{tt} \,  \mch
\Bigg[  y^R_{H^+tb}m_b (C_0+C_1+C_2)
+y^L_{H^+tb} m_t(C_1+C_2) \Bigg]
\,, \\ \nn
{\cal M}_{3}^{\rm (c)} &=&  -{\cal M}_{3}^{\rm (b)}
\Big( y^L_{H^+tb} \leftrightarrow  y^R_{H^+tb}\,, t \leftrightarrow b \Big)\,.
\end{eqnarray}
Here $C_l$ and $C_{lm} $  are
\begin{eqnarray}
C_{l,lm} &=& C_{l,lm}(m_W^2, 0, M_{H^+}^2, m_{b}^2, m_{t}^2, m_{t}^2)\,.
\end{eqnarray}

For $H^+ \to W^+ Z$, ${\cal M}_1$ is not related with ${\cal M}_2$,
given by
\begin{eqnarray}
\label{eq:FormFactor}
{\cal M}_{1}^{\rm (a)} &=& - \frac{ \gh^Z_{WW} }{\mch}
\Big(1-\frac{m_Z^2}{m_W^2}\Big) \left[
 y^R_{H^+tb} m_b \lf B_0 + B_1+\frac{1}{\epsilon}\ri
+y^L_{H^+tb} m_t \lf B_1-\frac{1}{\epsilon}\ri
 \right]\,, \\ \nn
{\cal M}_{1}^{\rm (b)} &=& \frac{y^R_{H^+tb} m_b}{M_{H^+}}\bigg[
\gh^L_{Ztt} m_W^2 \Big(C_0+3C_1+C_2+2C_{11}+2C_{12}\Big)
\\ \nn && \qquad \qquad \quad
- \gh^L_{Ztt} m_Z^2 \Big(C_0+C_1+C_2+2C_{12}\Big)
\\ \nn  &&\qquad \qquad \quad
-\gh^L_{Ztt} \left\{ 1-4C_{00} - M_{H^+}^2(C_0+C_1+3C_2 + 2C_{12} +2C_{22})\right\}
\\ \nn  &&\qquad \qquad \quad
-2\gh^R_{Ztt} m_t^2 C_0 + \frac{\gh^L_{Ztt}}{\epsilon}\bigg]
\\ \nn &&
+\frac{y^L_{H^+tb} m_t}{M_{H^+}}\bigg[
\gh^L_{Ztt} m_W^2 \Big(2C_1+C_2+2C_{11}+2C_{12}\Big)
-\gh^R_{Ztt} m_W^2 \Big(C_1+2C_{11}+2C_{12}\Big)
\\ \nn &&\qquad \qquad \quad
- \gh^L_{Ztt} m_Z^2 \Big(C_2+2C_{12}\Big)
+ \gh^R_{Ztt} m_Z^2 \Big(C_1+2C_{12}\Big)
\\  \nn  &&\qquad \qquad \quad
-\gh^L_{Ztt} \left\{ 1-4C_{00} - M_{H^+}^2(C_2 + 2C_{12} +2C_{22})\right\}
\\ \nn  &&\qquad \qquad \quad
+\gh^R_{Ztt} \left\{ 1-8C_{00} - M_{H^+}^2(C_1 +2C_2+ 2C_{12} +2C_{22})\right\}
\\ \nn &&\qquad \qquad \quad
+ \frac{(\gh^L_{Ztt} - 2\gh^R_{Ztt})}{\epsilon}
\bigg]
\,, \\ \nn
{\cal M}_{1}^{\rm (c)} &=& {\cal M}_{1}^{\rm (b)}
\Big( y^L_{H^+tb} \leftrightarrow  y^R_{H^+tb}\,, t \leftrightarrow b \Big)
\,.
\end{eqnarray}
The
${\cal M}_{2}$ and ${\cal M}_{3}$ for $H^\pm \to W^\pm Z$ are
\begin{eqnarray}
{\cal M}_{2}^{\rm (a)} &=&  0
\,, \\[3pt] \nn
{\cal M}_{2}^{\rm (b)} &=& -2y^R_{H^+tb}m_b \mch
\gh^L_{Ztt} (C_0+C_1+3C_2+2C_{12}+2C_{22})
\nn \\ \nn &&
+2y^L_{H^+tb}m_t \mch \Big[
-\gh^L_{Ztt} (C_2+2C_{12}+2C_{22})
+\gh^R_{Ztt} C_1
\Big]
\,,  \\[3pt] \nn
{\cal M}_{2}^{\rm (c)} &=&  {\cal M}_{2}^{\rm (b)}
\Big( y^L_{H^+tb} \leftrightarrow  y^R_{H^+tb}\,, t \leftrightarrow b \Big)
\,, \\[3pt] \nn
{\cal M}_{3}^{\rm (a)} &=& 0
\,,  \\[3pt] \nn
{\cal M}_{3}^{\rm (b)} &=& 2y^R_{H^+tb}m_b \mch
\gh^L_{Ztt} (C_0+C_1+C_2)
+2y^L_{H^+tb}m_t \mch\big(
\gh^L_{Ztt} C_2 +\gh^R_{Ztt} C_1\big)
\,, \\[3pt] \nn
{\cal M}_{3}^{\rm (c)} &=&  -{\cal M}_{3}^{\rm (b)}
\Big( y^L_{H^+tb} \leftrightarrow  y^R_{H^+tb}\,, t \leftrightarrow b \Big)\,.
\end{eqnarray}
The $B_l$ and $C_{lm} $ are as follows:
\begin{eqnarray}
B_l &=& B_l(M_{H^+}^2, m_{b}^2, m_{t}^2)\,,  \\ \nn
C_{lm} &=& C_{lm}(m_W^2, m_Z^2, M_{H^+}^2, m_{b}^2, m_{t}^2, m_{t}^2)\,.
\end{eqnarray}

\subsection{Decay Form Factors from the VLQ contributions}

We first present the form factors of ${\cal M}_{2}$ and ${\cal M}_{3}$ for $H^\pm \to W^\pm V$ ($V=\gm,Z$)
through the VLQ loop
as
\begin{eqnarray}
\label{eq:FormFactor:VLQ:M2}
{\cal M}_{2,ij}^{\rm (a)} &=&  0
\,, \\  \nn
{\cal M}_{2,ijk}^{\rm (b)} &=&
4 \gh^W_{\mcd_i \mcu_k} \gh^V_{\mcu_k \mcu_j} y^{H^+}_{\mcu_j \mcd_i} \mch
\\ \nn &&
\times \bigg[M_{\mcu_k}C_1
-M_{\mcd_i}\Big(C_0+C_1+3C_2+2C_{12}+2C_{22}\Big)
\\[3pt]  \nn&& \quad -M_{\mcu_j}\Big(C_2+2C_{12}+2C_{22}\Big)
\bigg]
\,,  \\ \nn
{\cal M}_{2,ijk}^{\rm (c)} &=& 4 \gh^W_{\mcd_k \mcu_i} \gh^V_{\mcd_j \mcd_k} y^{H^+}_{\mcu_i \mcd_j} \mch
\bigg[~ \mcu \leftrightarrow \mcd ~\bigg]
\,, \\  \label{eq:M3:VLF}
{\cal M}_{3,ij}^{\rm (a)} &=& {\cal M}_{3,ijk}^{\rm (b)} = {\cal M}_{3,ijk}^{\rm (c)} = 0
 \,,
\end{eqnarray}
where $\left[ ~\mcu \leftrightarrow \mcd ~\right]$ denotes interchanging $\mcu$
and $\mcd$ for the terms in the square parenthesis of the previous formula while remaining the indices.
Note that ${\cal M}_{3,ijk}^{\rm (b)} = {\cal M}_{3,ijk}^{\rm (c)} = 0$ because of the vectorlike nature of the VLFs,
i.e., $y^L_{H^+ \mcu\mcd} = y^R_{H^+ \mcu\mcd}$.
The full expressions of $B_l$, $C_l$, and $C_{lm} $ are
\begin{eqnarray}
B_l &=& B_l( \mch^2, M_{\mcd_i}^2, M_{\mcu_j}^2)\,, \nn \\
C_{l,lm} &=& C_{l,lm}(m_W^2, m_V^2, \mch^2, M_{\mcd_i}^2, M_{\mcu_k}^2, M_{\mcu_j}^2)\,.
\end{eqnarray}

For $H^\pm \to W^\pm Z$, ${\cal M}_1$'s are independent from ${\cal M}_2$, given by
\begin{eqnarray}
\label{eq:FormFactor:VLQ:M1}
{\cal M}_{1,ij}^{\rm (a)} &=& 2 \gh^W_{\mcd_i \mcu_j} \gh^Z_{WW}  y^{H^+}_{\mcu_j\mcd_i}\frac{1}{\mch}
\lf 1-\frac{m_Z^2}{m_W^2}\ri   \\ \nn
&& \times \left[ -2 M_{\mcd_i} B_0 - 2 (M_{\mcd_i}+M_{\mcu_j})B_1
+ \frac{1}{\epsilon}(M_{\mcu_j} - M_{\mcd_i}) \right]\,, \\  [3pt] \nn
{\cal M}_{1,ijk}^{\rm (b)} &=& 2 \gh^W_{\mcd_i \mcu_k} \gh^Z_{\mcu_k \mcu_j}  y^{H^+}_{\mcu_j\mcd_i} \frac{1}{M_{H^+}}
\\ \nn  &&
\times \bigg[
 M_{\mcd_i}m_W^2\lf C_0+3C_1+C_2+2C_{11}+2C_{12}\ri
-M_{\mcd_i}m_Z^2 \lf C_0+C_1+C_2+2C_{12} \ri
\\[3pt] \nn && \qquad
+M_{\mcd_j}m_W^2\lf 2C_1+C_2+2C_{11}+2C_{12}\ri
-M_{\mcd_j}m_Z^2\lf C_2+2C_{12}\ri
\\[3pt] \nn && \qquad
-M_{\mcd_k}m_W^2\lf C_1+2C_{11}+2C_{12}\ri
+ M_{\mcd_k}m_Z^2\lf C_1+2C_{12}\ri
\\[3pt] \nn  &&\qquad
-M_{\mcd_i}\left\{ 1-4C_{00} -\mch^2(C_0+C_1+3C_2 + 2C_{12} +2C_{22})\right\}
\\[3pt] \nn &&\qquad
- M_{\mcu_j}\left\{ 1-4C_{00} - \mch^2(C_2 + 2C_{12} +2C_{22})\right\}
\\[3pt] \nn &&\qquad
+M_{\mcu_k} \left\{ 1-8C_{00} - \mch^2(C_1 +2C_2 + 2C_{12} +2C_{22})\right\}
\\[3pt] \nn  &&\qquad
-2 M_{\mcd_i } M_{\mcu_j} M_{\mcu_k} C_0
+ \frac{1}{\epsilon} (M_{\mcd_i} + M_{\mcu_j} - 2M_{\mcu_k})
\bigg]
\,,\\ \nn
{\cal M}_{1,ijk}^{\rm (c)} &=& 2 \gh^W_{\mcd_k \mcu_i} \gh^Z_{\mcd_j \mcd_k}  y^{H^+}_{\mcu_i\mcd_j} \frac{1}{M_{H^+}}\bigg[~ \mcu \leftrightarrow \mcd ~\bigg]
\,.
\end{eqnarray}

\end{document}